\documentclass[a4paper,12pt]{article}

\usepackage{lmodern}
\usepackage[T1]{fontenc}
\usepackage{mathtools}
\usepackage{amsmath}
\usepackage{amssymb}
\usepackage{multicol}% http://ctan.org/pkg/multicols
\usepackage{graphicx}
\usepackage[font=footnotesize]{caption}
\usepackage{subcaption}
\usepackage{anysize} % Soporte para el comando marginsize
\marginsize{2.5cm}{2.5cm}{1cm}{1cm}
\graphicspath{{graficas/}}
\usepackage{float}
\usepackage{xcolor}

\usepackage{soul}
\usepackage{cite}
\usepackage[super]{nth}

\begin{document}

\title{Lotka-Volterra systems with stochastic resetting}

\author{Gabriel Mercado-V\'asquez and Denis Boyer\footnote{\texttt boyer@fisica.unam.mx}\\
{\normalsize\it Instituto de F\'isica, Universidad Nacional Aut\'onoma de M\'exico, Mexico City 04510, Mexico}}

%\begin{document}
\maketitle

    \begin{abstract}
    We study the dynamics of predator-prey systems where prey are confined to a single region of space and where predators move randomly according to a power-law (L\'evy) dispersal kernel. Site fidelity, an important feature of animal behaviour, is incorporated in the model through a stochastic resetting dynamics of the predators to the prey patch. We solve in the long time limit the rate equations of Lotka-Volterra type that describe the evolution of the two species densities. Fixing the demographic parameters and the L\'evy exponent, the total population of predators can be maximized for a certain value of the resetting rate. This optimal value achieves a compromise between over-exploitation and under-utilization of the habitat. Similarly, at fixed resetting rate, there exists a L\'evy exponent which is optimal regarding predator abundance. These findings are supported by 2D stochastic simulations and show that the combined effects of diffusion and resetting can broadly extend the region of species coexistence in ecosystems facing resources scarcity.        
    \end{abstract}
    
%\maketitle

\section{Introduction}
	In situations of rapid population growth, species that exploit limited resources can face scarcity and risks of extinction. An essential question in ecology and population dynamics is to identify the strategies that individuals may adopt in unfavourable conditions for their own survival, as well as for the persistence of large populations. Among these strategies, mobility plays an important role. In recent years, it has been recognized that population dynamics is strongly affected by the spatial structure of the environment \cite{Ilkka1998} and by the way living organisms move through it \cite{Turchin98,nathan2008movement}. For instance, species abundances depend on the spatial distribution of prey patches and how predators commute between these \cite{metapopulationDiekmann,holland2008strong}. In experiments of bacterial predator-prey systems, species may coexist thanks to diffusion and a non-uniform spatial distribution of individuals \cite{huffaker1958experimental,mcmurtrie1978persistence}. 
%In this context spatial models have gained relevance given that they represent a better approximation to many real situations than low dimensional dynamical systems based on ordinary differential equations \cite{VolterraVariations,LotkaElements}. 

Fragile ecosystems, {\it i.e.}, systems close to extinction thresholds, are often spatially fragmented or composed of small populations well separated in space \cite{metapopulationDiekmann}. For this reason, single patch systems where resources or prey are restricted to a limited area have attracted a particular interest from both practical and theoretical point of views \cite{populationbiologyhastings}.
%[buscar otras referencias single patch y resumir algunos resultados]. 
The one-patch configuration, a limiting case of a fragile system, allows to study in details the effects of the predator mobility on species survival within and outside the patch \cite{metapopulationDiekmann,maynard1978models}.

In spatially explicit population models, animal movements are usually modeled by Brownian diffusion, a mode of transport that reflects the dispersion of individuals performing ordinary Markovian random walks \cite{solebascompte,Mobilia}. In situations of low prey density, however, predator species often forage under conditions of high uncertainty and move across many landscape scales to search for resources \cite{bartumeus2016foraging}. On the other hand, when they have located profitable areas, foragers often exhibit site fidelity \cite{switzer1993site}. The random walk description obviously ignores these two important features of animal mobility, namely, multiple-scale movements and memory effects. 

In recent years, technological advances in individual tracking have revealed that animal foraging movements are often better described by L\'evy flights or L\'evy walks than ordinary random walks \cite{gandhi1999,gandhibook, bartu2003,RF,boyermonos,reynolds,Brown,Sims,deJager,
humphries}. L\'evy flights are random walks composed of independent displacement steps, whose lengths $l_i$ are broadly distributed, following a power-law distribution with infinite variance, or $p(l)\sim l^{-\beta}$ at large $l$, with $1<\beta<3$ \cite{chechkin2008introduction}. Under environmental uncertainty, multi-scale movement represents a mechanism by which exploratory behaviour and information gathering can be improved: L\'evy flights allow an increase in the number of visited sites, without preventing returns to previous sites\cite{bartumeus2016foraging}. In this context, the anomalous diffusion generated by L\'evy motion can represent an efficient random search strategy for finding prey items in unpredictable environments, when prey are scarce and distributed in patches \cite{gandhi1999,gandhibook}.       

Although the L\'evy flight model has been extensively applied to animal movement in recent years, other paradigms exist. We mention for instance that the long range migration of individuals on a lattice is sometimes modeled by small world Watts-Strogatz networks, where quenched short-cuts connect distant lattice sites. The lengths of these short-cuts can be chosen to follow a power-law distribution, in analogy with L\'evy diffusion\cite{Jespersen2000,Kozma2007}. Such network model exhibits diffusive behaviours similar to Levy flights for $\beta>2$, as far as the probability of presence at the origin or the number of distinct visited sites is concerned. Nevertheless, differences with Levy flights are noticeable in 1D when the exponent parameter $\beta$ is smaller than 2, where the processes are transient.

Long range displacements can have a drastic impact on ecosystems dynamics. In cyclic Lotka-Volterra (LV) models such as the rock-scissor-paper games, the addition of long range links to a regular lattice can give rise to a transition to global oscillations\cite{Szabo2004,Shabunin2008}. In such systems without strict competitive hierarchy, local mobility is nevertheless advantageous for biodiversity: extinctions can occur if the density of long range links is too high\cite{Szabo2004}. Similarly, diversity can be lost in similar models if the mobility of individuals is too large\cite{kerr2002local,reichenbach2007mobility}.

Many animal species also rely on spatial memory during foraging \cite{fagan2013spatial} and individuals often exhibit a tendency to regularly revisit places that are associated with successful foraging experiences \cite{thomson1997trapline,EcologyThomas1982}. This behaviour can occur for instance during trap-lining, which consists in visiting several known resource sites before returning to a shelter, or by cropping, when foragers exploit a particular area for some time and then temporarily abandon it to give the resources the time to regenerate \cite{champanAnimal1990}. A simple way to model fidelity to a profitable patch consists in interrupting diffusion stochastically (say, at some rate $r$) and to reset the forager position to the resource patch. From this site, a new diffusing path is then generated until the next resetting event. This model, of interest in the present work, is motivated by the fact that the properties of ordinary diffusing particles subjected to stochastic resetting are mathematically tractable, as shown by a body of recent studies \cite{PhysRevLett.106.160601,evans2011diffusion,mendezcampos2016,pal2016}, including extensions to the case of L\'evy flights \cite{kusmierz2014first,kusmierz2015optimal}. An important effect of resetting to a single site on diffusion is to generate at large time a stationary probability density profile around the resetting point, in similitude with the home range behaviour of animals \cite{borger2008there} and in contrast with the vanishing densities of unbounded diffusing processes. For a particle that diffuses in an ordinary way and is subject to resetting, the mean square displacement reaches a finite asymptotic value. 
This type of motion was shown to represent an efficient search strategy to find hidden targets by a single searcher, but was not investigated so far in population models.
%: the mean time when the particle (starting at some origin) hits a fixed target position for the first time can be minimized for a certain value of $r$ \cite{PhysRevLett.106.160601}.

Throughout this paper, we define optimal strategies as the ones that maximize the abundance of a predator population (although other related quantities can be considered). This point of view differs from the usual definitions of efficiency in random search problems with a single forager, which typically focus on the individual foraging efficiency through the mean first hitting time to a target. There is in principle an important distinction between the reproductive interest of an individual and the survival of a species, as noticed by Williams some time ago\cite{williams1966natural}. At the population scale, because of interactions and demographic constraints, individually efficient foragers do not necessarily have a positive impact on abundances. For instance, efficient predators may over-exploit resources, leading to low abundance and even extinctions. Conversely, less efficient foragers may give resources the time to regenerate, allowing that same forager species to survive over many generations.

In a recent work, we studied a model of the Lotka-Volterra type, where a colony of mortal predators performing ordinary random walks or L\'evy flights fed on many small prey patches separated in space\cite{pnas}. The predators had a fixed death rate, could diffuse freely outside the patches and could only reproduce within a patch. It was shown that the use of L\'evy strategies could maximizes forager abundances and avoid extinctions, which were likely in the Brownian regime. Here, we continue with the methodology used in\cite{pnas} and consider the case of prey that are confined to a single patch and predators that are subjected to stochastic resetting to this patch, which models memory effects. The purpose of modifying the previous model is to investigate how (anomalous) diffusion and site fidelity combined together impact species coexistence. When the predator reproductive rate is low, we expect that in the absence of resetting the fluxes of diffusing foragers leaving the patch will not be compensated by the birth of new foragers from individuals that are still in the patch. In those cases, low abundances or even extinctions are likely. In the same conditions, a large resetting rate will sharply localize the foragers around the patch, increasing their density as well as competition for prey, which may negatively affect the overall abundance. One hence expects that the predator population reaches a maximum at some finite resetting rate. Similarly, for a fixed resetting rate, the predator abundance may be maximized by tunning the L\'evy exponent $\beta$ to a certain value. We investigate more generally how the movement parameters $r$ and $\beta$ that maximize predator abundances depend on the reproduction and death rates.

%The referee is right to point out this important aspect that was maybe not sufficiently discussed in the first version. A first point is that we define here optimal strategies in terms of populations rather than in terms of individual fitness (capture efficiency), the latter being common in single forager random search problems. We now sustain in the revised version our approach by mentioning the concepts advanced by Williams (1966) about the distinction between "organic adaptation" and "biotic adaptation", that is basically the debate of "individual interest" versus "population survival" when dealing with population dynamics. Individually efficient foragers (for instance, those that return often to the patch or with high resetting rates) may have negative collective consequences, such as overexploitation and low abundance for the populations. Conversely, less efficient searchers may allow population to survive over many generations. These points are now clarified in the Introduction. A second point is that, from a collective point of view, a natural quantity to measure the efficiency of a strategy is the predator abundance. In the deterministic approach that we use, this quantity evolves toward a stable steady state solution (there are no oscillations in our model). In a stochastic approach, however, there are other ways of quantifying the efficiency, such as those proposed by the referee. We did not focus on stochasticity in this manuscript, although it is an important aspect deserving future research. We have mentioned this issue in the Conclusion.
	
\section{Analytical Lotka-Volterra model with patchy prey and mobile predators}
	
We re-take here the setup of the model presented in \cite{pnas}. Let us consider a regular D-dimensional lattice of square cells with length $R$, where the positions of the center of each cell is represented by a vector $\mathbf{n}$ with integer components. Predators occupy the cells and perform independent random motion (the resetting part will be introduced further below). The lengths of these displacements are in principle continuous and we set the unit length as the smallest possible displacement performed by a predator ($1<R$). We denote the probability distribution of the dispersal of a predator between cells as:  
	\begin{equation}
	p\left(\textbf{\textit{l}}\right)=p_0\delta_{\textbf{\textit{l}},\textbf{0}}+\left(1-p_0\right)f\left(\textbf{\textit{l}}\right),
	\label{pd}
	\end{equation}
where $\textbf{\textit{l}}$ is a vector with integer components.	
The quantity $p_0$ represents the probability that the predator remains in the same cell after moving. This is possible because each cell has a spatial extent $R$ and a predator may not exit the cell ($\textbf{\textit{l}}=\mathbf{0}$) if its actual displacement is too small (see further below Eq. \ref{probabilityzero}). The function $f(\textit{\textbf{l}})$ is a normalized dispersal distribution between non-identical cells and will take the form of an inverse power law with exponent $\beta$ in the following, more detail being given in Section 2.2. It satisfies $f(\mathbf{0})=0$ and the normalization condition:
	
	\begin{equation}
		\sum_{\textit{\textbf{l}},\textbf{|\textit{l}|$\neq$0}} f(\textbf{\textit{l}})=1
	\end{equation}
	
    In this model, only the origin cell $\mathbf{n}=(0,0)$ can contain prey and it is also the only place where predators reproduce. Assuming that the number of individuals per cell is large, we neglect fluctuations and write two deterministic rate equations of the Lotka-Volterra type that are inspired of the model studied in \cite{Mobilia}. One equation describes the time evolution of the density of predators in cell $\textbf{n}$ at time $t$, denoted as $a(\textbf{n},t)$, and the other, of the density of prey $b(t)$ at the origin cell (the prey patch). As mentioned earlier, the predators can also relocate (reset) stochastically to the origin cell, independently of their current position in the system. This process happens at rate $r$. Once the predator has reset to the origin, it continues its dynamics from there, until the next resetting event in a way analogous to the problem studied in \cite{PhysRevLett.106.160601}. The equation for the predator density reads:

	\begin{equation}	
	\begin{aligned}
		\frac{\partial{a(\textbf{n},t)}}{\partial{t}}=&
		-\overbrace{\alpha(1-p_0)a(\textbf{n},t)}^{\nth{1} \text{ term}}+\overbrace{\alpha\sum_{\textit{\textbf{l}},|\textbf{\textit{l}}|\neq \textbf{0}}p(\textbf{\textit{l}})a(\textbf{n}-\textbf{\textit{l}},t)}^{\nth{2} \text{ term}}\\
		&+\underbrace{\lambda a_0(t)b(t)\delta_{\textbf{n,0}}}_{\nth{3} \text{ term}}-\underbrace{\mu a(\textbf{n},t)}_{\nth{4} \text{ term}}-\underbrace{ra(\textbf{n},t)}_{\nth{5} \text{ term}}+\underbrace{\left[r\sum_{m}a(\textbf{m},t)\right]\delta_{\textbf{n},0}}_{\nth{6} \text{ term}}
	\end{aligned}
		\label{lotka}
	\end{equation}
	
	The first term in the right-hand-side represents the rate at which predators leave the cell $\textbf{n}$, and $\alpha$ is the movement rate. The second term describes the rate at which predators arrive at the cell from another one through a random step, the third term accounts for reproduction at rate $\lambda$ (which only occurs at the origin cell, where prey are present), and the fourth term represents predator mortality, where $\mu$ is the mortality rate. The last two terms describe the resetting process and depart from the model of \cite{pnas}: the fifth term indicates that predators are removed from their current location at rate $r$, while the last term represents the increase of the density at the origin due to predators reset from everywhere in the system. It is clear that new predators are not created at the origin, they are just brought from elsewhere at rate $r$.
	\smallskip
	
		The prey density $b(t)$ in the prey patch obeys the ordinary differential equation:
	
	\begin{equation}
	\frac{db}{dt}=\sigma b\left(1-\frac{b+a_0}{K}\right)-\lambda' a_0 b
	\label{preydensity}
	\end{equation}
	where we have re-noted $a_0(t)=a(0,t)$ as the predator density at the prey patch, $\sigma$ being the prey reproductive rate and $\lambda'$ the predation rate. $K$ represents the prey carrying capacity, which, as is usual in population biology modelling\cite{Mobilia,populationbiologyhastings}, enforces the fact that prey cannot grow indefinitely but are limited by finite resources instead. Note that this realistic assumption was not present in the original Lotka-Volterra model \cite{LotkaElements,VolterraVariations}. Hence, in the absence of predators, the prey density asymptotically reaches its maximum value $K$.
	\smallskip

In the absence of movement ($\alpha=0$, $r=0$), populations vanish everywhere at large time except at the prey patch, where equations (\ref{lotka})-(\ref{preydensity}) reduce to two ordinary differential equations which have been studied in detail in Ref. \cite{Mobilia}. They admit two simple stationary fixed points, $(a_0^{(o)},b_0^{(o)})=(0,0)$ and $(a_0^{(u)},b_0^{(u)})=(0,K)$, corresponding to total extinction (over-exploitation point) and predator extinction (under-exploitation point), respectively. A third, globally stable coexistence fixed point exists for $\mu/\lambda<K$, given by
$a^{(coex)}_0=(K-\mu/\lambda)/(1+\lambda'K/\sigma)$
and $b^{(coex)}_0=\mu/\lambda$.  (If $K<K_c=\mu/\lambda$, predators go extinct and $b^{(coex)}_0=K$.) Oscillatory solutions do not exist in this problem \cite{Mobilia} and this feature probably holds in the space-dependent problem as well. See \cite{TauberPopul} for an extended discussion on this latter aspect.

\subsection{Stationary solution}
In all the following we focus our analysis on steady state solutions. Expressions for the stationary prey and predator densities, denoted by $a_\textbf{n}$ and $b$, respectively, can be derived by setting the time derivatives to zero. For the prey, one obtains the trivial solution $b=0$ and
	
	\begin{equation}
		b=K-a_0(1+K\lambda'/\sigma),
		\label{prey}
	\end{equation}
where $a_0$ is the stationary predator density at the prey patch, a constant yet to be determined. We introduce the discrete Fourier transforms:
	
	\begin{equation}
		\hat{a}(\textbf{k})\equiv\sum_{\textbf{n}}a_{\textbf{n}}e^{-i\textbf{k}\cdot\textbf{n}},\quad \hat{f}(\textbf{k})\equiv\sum_{\textit{\textbf{l}}} f(\textit{\textbf{l}})e^{-i\textbf{k}\cdot\textit{\textbf{l}}},
	\end{equation}
and take the Fourier transform of Eq. (\ref{lotka}) with the steady state condition. One obtains:
		
	\begin{equation}
	\hat{a}(\mathbf{k})=\frac{\lambda a_0\left[K-a_0(1+K\lambda'/\sigma)\right]+rA}{	\alpha\left(1-p_0\right)\left[1-\hat{f}(\textbf{k})\right]+\mu+r},
	\label{afourier}
	\end{equation}
where $A=\sum_{\textbf{m}}a_{\textbf{m}}=\hat{a}(\textbf{k}=\textbf{0})$. This solution differs from the predator density calculated in\cite{pnas}, due to the presence of resetting and the single patch geometry. By setting $\textbf{k}=\textbf{0}$ in Eq. (\ref{afourier}) we can solve for $A$  (note that $\hat{f}(\textbf{k}=\textbf{0})=1$ by normalization):
	
	\begin{equation}
	A=\frac{\lambda a_0}{\mu}\left[K-a_0\left(1+K\lambda'/\sigma\right)\right]
	\label{Azero}
	\end{equation}
A quantity of primary interest here is the total population of predators ($N_p$). It is simply given by:
	
	\begin{equation}
	N_p=R\sum_{\textbf{n}}a_{\textbf{n}}=R\hat{a}(\textbf{k}=\textbf{0})
	\label{totpopul}
	\end{equation}
Replacing into Eq. (\ref{totpopul}) the quantity $a(k=0)$ given by Eq. (\ref{Azero}) gives:
	
	\begin{equation}
		N_p=R\frac{\lambda a_0}{\mu}\left[K-a_0\left(1+K\lambda'/\sigma\right)\right]
		\label{totpopul2}
	\end{equation}	
The predator density in the prey patch $a_0$ can be obtained self-consistently from the inverse Fourier transform of $a(k)$ evaluated at the origin $n=0$. We use Eq. (\ref{afourier}) and the inverse relation 

 \begin{equation}
 	a_\mathbf{n}=\frac{1}{(2\pi)^D}\int_{\cal{B}}\hat{a}(\mathbf{k})e^{i\mathbf{n}\cdot\mathbf{k}}d\mathbf{k},
 	\label{inverse}
 \end{equation}
 where $\cal{B}$ means that we are integrating over the Brillouin zone, which in one dimension (D=1) is the interval $-\pi<k<\pi$, and in two dimensions (D=2), the domain $-\pi<k_x,k_y<\pi$. Setting $\textbf{n}=\textbf{0}$ we obtain an equation for $a_0$, which admits the trivial solution $a_0=0$ and a non-trivial one:
	
	\begin{equation}
 a_0=\frac{1}{1+K\lambda'/\sigma}\left\{K-\left[\frac{\lambda}{\mu(2\pi)^D} \int_{B}\frac{(\mu+r)d\textbf{k}}{\alpha(1-p_0)[1-\hat{f}(\textbf{k})]+\mu+r}\right]^{-1}\right\}
	\label{azero}
	\end{equation} 
	\smallskip
If the above solution happens to be negative, then only the solution $a_0=0$ is acceptable.  
\smallskip

	Eqs. (\ref{totpopul2}) and (\ref{azero}) fully determine the total number of predators in the system as a function of the different parameters. The number of prey is deduced from Eq. (\ref{prey}). We notice from (\ref{totpopul2}) that $N_p$ obeys a logistic relation with respect to $a_0$, the predator density at the origin patch. Therefore, the total number of predators is maximal when $a_0=a_0^{max}\equiv K/[2(1+K\lambda'/\sigma)]$ whereas $N_p$ vanishes at $a_0\equiv 0$ and at $a_0 \equiv K/(1+K\lambda'/\sigma)=2a_0^{max}$.
	\smallskip
	
 Eq. (\ref{totpopul}) shows that in the low density regime, {\it i.e.}, $0<a_0<a^{max}_0$, an increase in $a_0$ produces an increase of $N_p$: this is the under-exploitation regime. Whereas for $a^{max}_0<a_0<2a^{max}_0$, the high density regime, any increase in $a_0$ decreases the total population (over-exploitation regime). When  $a_0>2a_0^{max}$, the solution is no longer valid and the only non-negative steady state solution of Eq.(\ref{preydensity}) is $b=0$, implying $a_0=0$ and $N_p=0$. This point corresponds to the global extinction of prey and predators.
	\smallskip

	Using Eq. (\ref{azero}) and solving the equation $a_0=a_0^{max}$, we obtain a relation (which may not always admit solutions) between the movement parameters $r$ and $\beta$ and the demographic parameters, and such that the predator abundance in the whole system is maximal. This relation reads: 
	
	\begin{equation}
	\frac{1}{(2\pi)^D}\int_{B}\frac{d\textbf{k}}{(1-p_0)[1-\hat{f}(\textbf{k})]+\mu^*+r^*}=\frac{2\mu^*}{(r^*+\mu^*)\lambda^* K},
	\label{findmaximum}
	\end{equation}
	\smallskip
where the reduced parameters are defined as $\mu^*=\mu/\alpha$, $r^*=r/\alpha$ and $\lambda^*=\lambda/\alpha$. To simplify the notation, from now on we will set the movement rate $\alpha$ to $1$ and write the reduced parameters without an asterisk. Thus, fixing the ecological parameters $\mu$, $\lambda$
and $K$, there exists in principle a set of pairs $(\beta,r)$ that satisfy Eq. (\ref{findmaximum}). We recall that the dependence in $\beta$ is contained in $\hat{f}(\textbf{k})$.

    To gain insights on the stationary states calculated above, we will consider in the numerical applications the simpler $1D$ case (results in $2D$ are qualitatively similar, as will be shown further below). The jump probability distribution in Eq. (\ref{pd}) becomes:
	
		\begin{equation}
	p\left(\textit{l}\right)=p_0\delta_{\textit{l},\textit{0}}+\left(1-p_0\right)f\left(\textit{l}\right)
	\label{pdonedimension}
	\end{equation}
	where \textit{l} = $0$, $\pm 1$, $\pm 2$, etc.
	
	\subsection{Predator movements}
	
	We now calculate the probability $p_0$ for an agent to leave a cell of length $R$ in the one dimensional case. The same calculation can be found in the Supplemental Information of\cite{pnas}.  We assume that predators perform continuous steps of length $x$, which are independent and identically distributed. We choose a symmetric, normalized power-law distribution:

	\begin{equation}
	\psi(x)=
	\frac{\beta-1}{2}|x|^{-\beta} \quad \text{for} \quad  |x|>1
    \label{psix}
	\end{equation} 
	
	The exponent $\beta$ characterizes the forager displacements, which have a minimal length of 1, which means that $\psi(x)=0$ for $|x|<1$. The three important movement scenarios are: 1) When $\beta$ $\geq$ $3$, the predators diffuse normally similarly to Brownian motion. 2) If $1<\beta<3$, the movements correspond to L\'evy flights. 3) In the highly super-diffusive regime $\beta \approx 1$, the foragers take extremely long steps.
	\smallskip
	
	 We assume that the walkers occupying the patch can be anywhere inside with a uniform probability. The average probability that a step does not bring the walker out of the cell is:
	
	\begin{equation}
	p_0=\frac{2}{R}\int_{0}^{R} dr\int_{0}^{r} dx\  \psi(x)
	\label{probabilityzero}
	\end{equation}
	The factor 2 takes into account the right and left moves. One obtains:	
	\begin{equation}
	p_0=
	1-\frac{1}{R}+\frac{1-R^{2-\beta}}{(2-\beta)R}  \quad {\rm for} \quad \beta\neq 2
	\end{equation}	
and $p_0= 1-\frac{1}{R}-\frac{\ln(R)}{R}$  for $\beta=2$. When $\beta$ is large (Brownian diffusion), $p_0\simeq 1-2/R\longrightarrow 1$ at large $R$: only the predators located close to the boundary of the cell can exit the cell in one step. Conversely $p_0\longrightarrow0$ as $\beta\longrightarrow1$, as is expected if the walkers take very long steps.
	Following the same arguments, one may calculate the probability that the walker jump over $l$ cells in one step, with $l$ a non-zero integer. For simplicity, we assume for this quantity a normalized power-law distribution with the same exponent $\beta$ as in Eq. (\ref{psix}):
	
	\begin{equation}
	f(\textit{l})=\frac{|l|^{-\beta}}{2\sum_{m=1}^{\infty} m^{-\beta}}\quad \textrm{  for  }l=\pm 1, \pm 2, ...	
	\end{equation}

	\section{Results in 1D}
	
	Without lack of generality, we fix $K=1$ and $\sigma=1$ in the following, since these parameters appear only as prefactors of the predator abundance $N_p$, see Eqs. (\ref{totpopul2}) and (\ref{azero}). The size to the prey patch is set to $R=10$. 

 \subsection{Predator populations}
 
We first focus on the variations of the total number of predators with respect to the mortality rate $\mu$ and the exponent $\beta$, for various values of the resetting rate and of the reproduction rate $\lambda$. The expressions are calculated using Mathematica.
	\smallskip
	
\begin{figure}[!htp]
		\centering
		\begin{subfigure}{0.3\textwidth}
			\includegraphics[width=\textwidth]{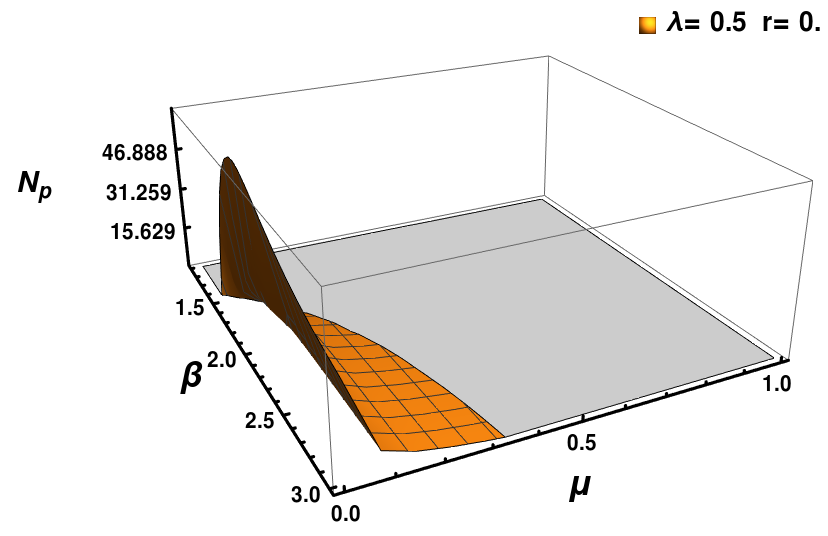}
			\caption{$r=0.0$}
			\label{fig:btamua}
		\end{subfigure}
		\hfill
		\begin{subfigure}{0.3\textwidth}
			\includegraphics[width=\textwidth]{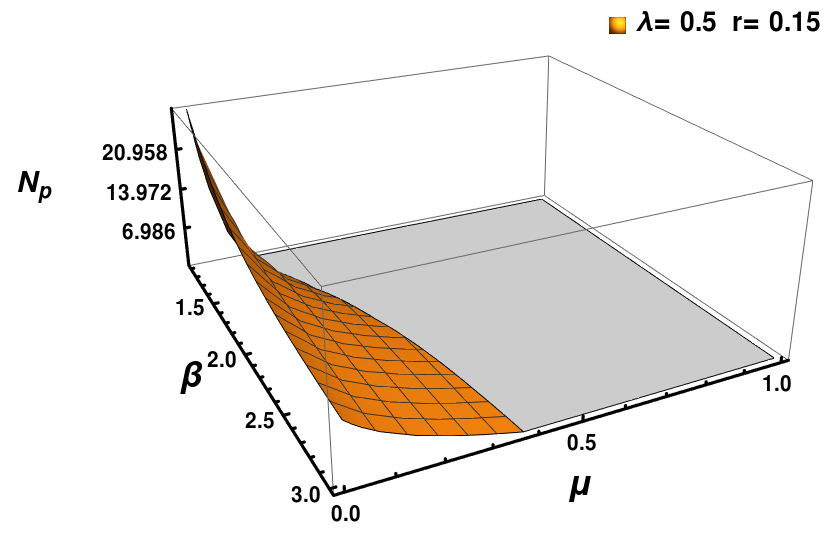}
			\caption{$r=0.15$}
			\label{fig:btamub}	
		\end{subfigure}
		\hfill
		\begin{subfigure}{0.3\textwidth}
			\includegraphics[width=\textwidth]{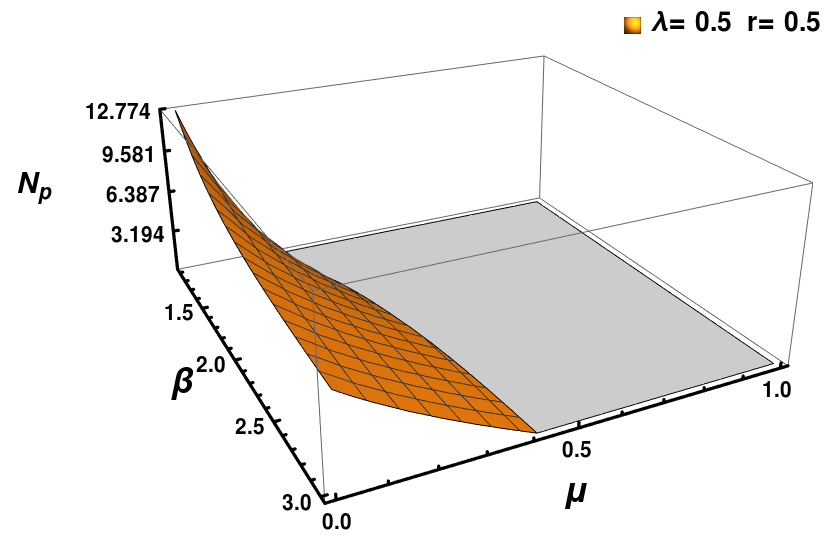}
			\caption{$r=0.5$}
			\label{fig:btamuc}
		\end{subfigure}
		\caption{Total population of predators in the ($\mu$, $\beta$)-plane for three values of $r$ and $\lambda=0.5$.}
		\label{btamu}
	\end{figure}

	Figure \ref{btamu} represents the case of a system with $\lambda=0.5$. At $r=0$ (no resetting, or free diffusion) and at fixed $\mu$, predators become extinct below a certain value of $\beta$, which increases as $\mu$ increases (see Fig. \ref{fig:btamua}). This means that if predators that are too super-diffusive they escape from the patch and do not return to it for predation and reproduction, leading to extinction. At high mortality rates, even Brownian predators ($\beta>3$) cannot avoid extinction. On the other hand, at low mortality rates, $N_p$ takes a maximum value when $\beta\approx 2$, in the L\'evy regime. 
	\smallskip

	If $r>0$ (Fig. \ref{fig:btamub}-\ref{fig:btamuc}), the predators regularly return to the prey patch due to resetting, independently of their Levy exponent. At low $\mu$, predators tend to over-exploit the patch and $N_p$ is larger when $\beta$ tends to 1, corresponding to the highly super-diffusive regime. However, when $\mu$ takes large values we can find a behaviour similar to that found for the case of  $r=0$: For a fixed $\mu$, $N_p$ vanishes below a certain value of $\beta$. 
	In all cases,  above a critical $\mu$, predators go extinct due to large predators mortality. The critical value of $\mu$ increase as $r$ increases, see Figure \ref{fig:btamuc}.
	\smallskip
	
\begin{figure}[htbp]
	\centering
    \begin{subfigure}{0.3\textwidth}
		\includegraphics[width=\textwidth]{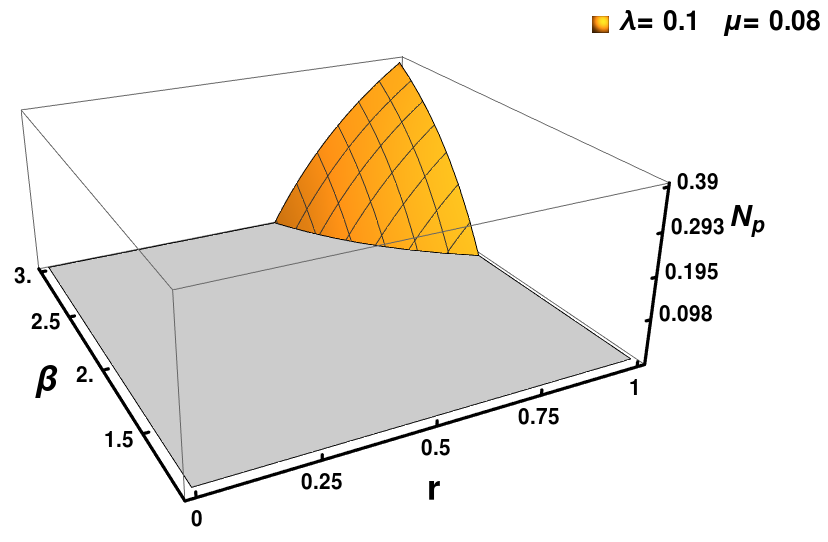}
		\caption{$\mu=0.08$}
		\label{fig:btaerec}	
	\end{subfigure}
	\hfill
	\begin{subfigure}{0.3\textwidth}
		\includegraphics[width=\textwidth]{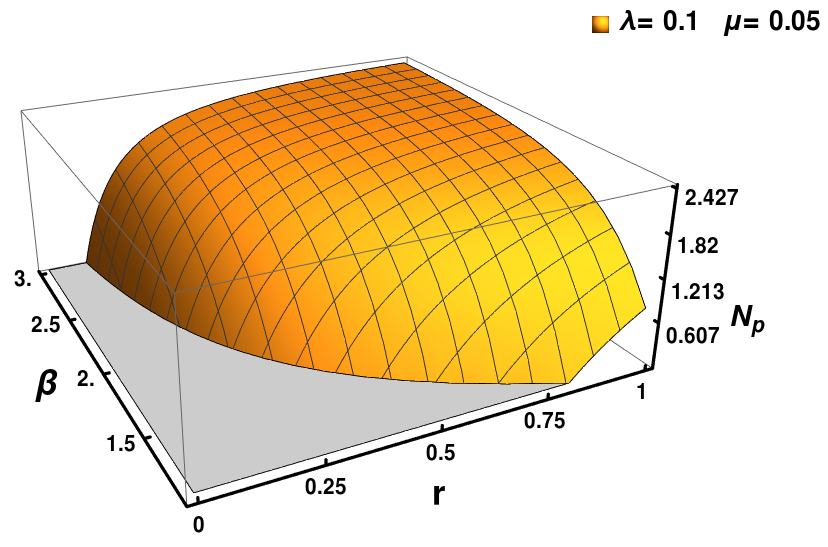}
		\caption{$\mu=0.05$}		
		\label{fig:btaereb}
	\end{subfigure}
    \hfill
	\begin{subfigure}{0.3\textwidth}
		\includegraphics[width=\textwidth]{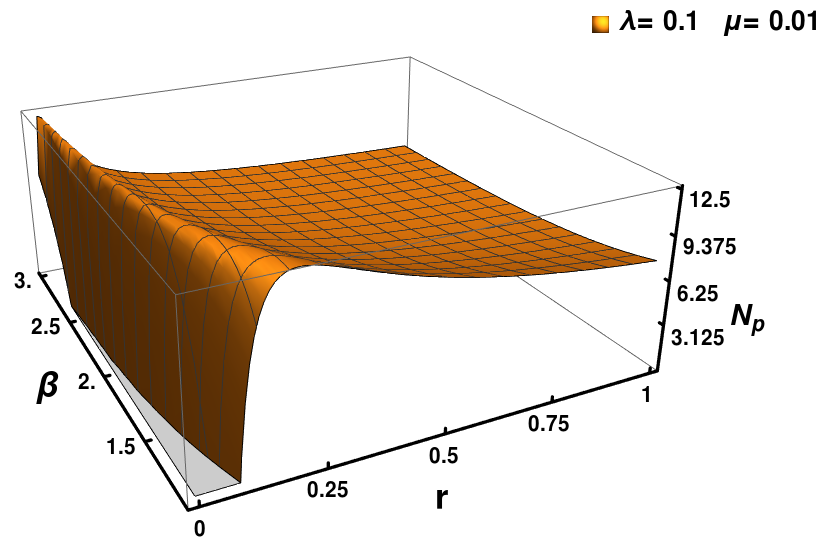}
		\caption{$\mu=0.01$}
		\label{fig:btaerea}		
	\end{subfigure}
    
    \caption{Total population of predators in the ($\beta$, $r$)-plane for three values of $\mu$ and $\lambda=0.1$.}\label{btaere}
\end{figure}

In Figure \ref{fig:btaerea}, the reproduction rate is lower than in Fig. \ref{btamu}. As shown by Fig. \ref{fig:btaerec}-b, where $\mu$ is relatively large (meaning that the conditions are unfavourable for predators), $N_p$ can become non-zero for $r>0$. At fixed $\beta$, $N_p$ increases with $r$, which indicates the positive effect of resetting on predator abundance. At fixed $r$, Brownian foragers (larger $\beta$) have larger populations.
\smallskip

When $\mu$ is smaller (Fig. \ref{fig:btaerea}), the behaviour is quite different: 1) the region with non-zero predator population is much larger, and 2) at fixed $\beta$, a maximal population abundance is reached for a finite optimal value of $r$. If $r$ is too small resources are underexploited and predator may go extinct like in the $r=0$ case, whereas larger values lead to resource overexploitation, which limits predator reproduction. Note that with $r$ fixed and varying $\beta$, the variations are much softer. The resetting process has stronger effects here on the populations than the type of movement between resetting events.
\smallskip

\begin{figure}[htbp]
		\centering
		\begin{subfigure}{0.3\textwidth}
			\includegraphics[width=\textwidth]{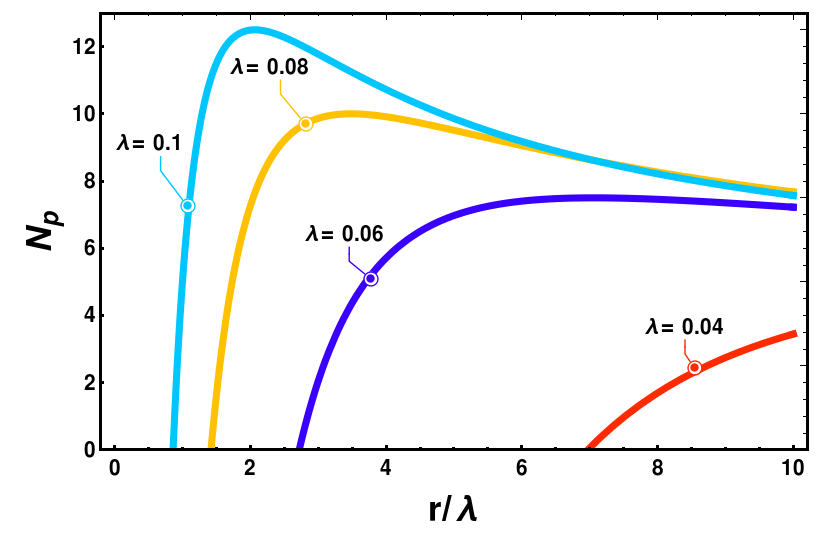}
			\caption{$\beta=1.1$}
			\label{varerea}
		\end{subfigure}
		\hfill
		\begin{subfigure}{0.3\textwidth}
			\includegraphics[width=\textwidth]{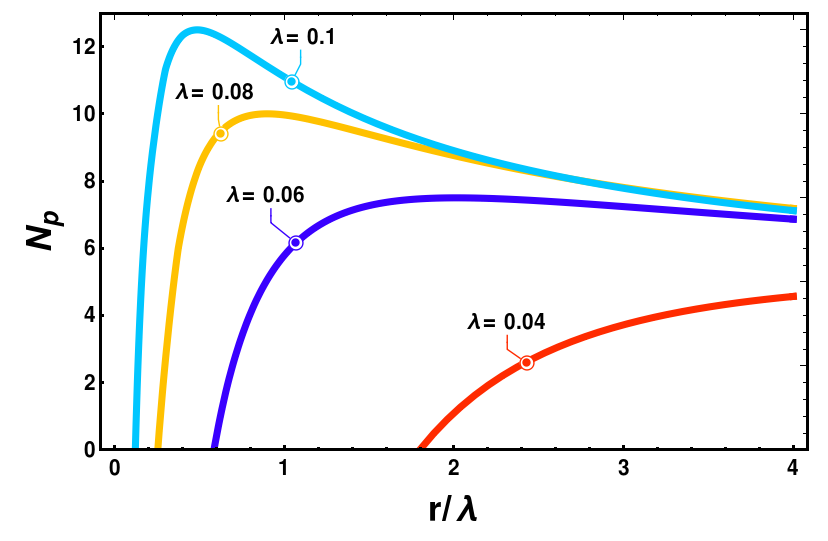}
			\caption{$\beta=2.0$}
			\label{varereb}
		\end{subfigure}
		\hfill
		\begin{subfigure}{0.3\textwidth}
			\includegraphics[width=\textwidth]{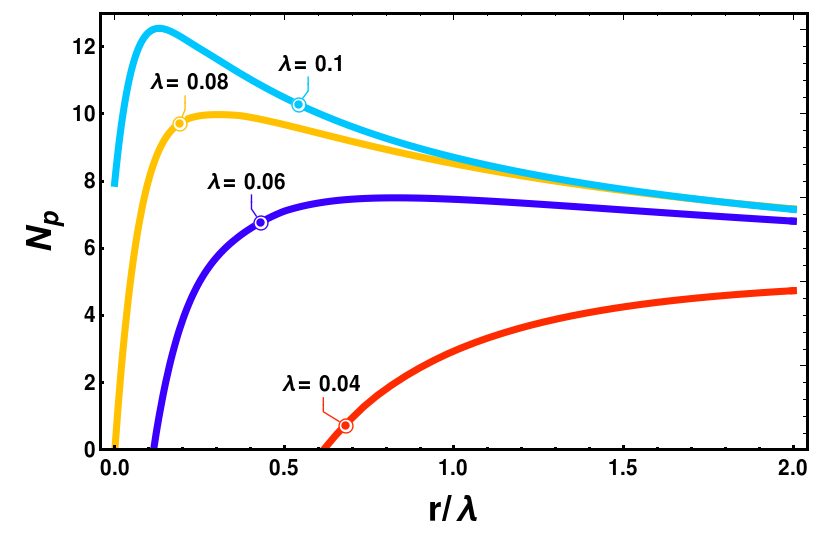}
			\caption{$\beta=3.0$}
			\label{varere}		
		\end{subfigure}
		\caption{Total population of predators as a function of the resetting rate, for several values of $\beta$ and high reproductive rates $\lambda$. (In all cases, $\mu=0.01$).}
		\label{variacionr}
	\end{figure}

\begin{figure}[htbp]
	\centering
	\begin{subfigure}{0.3\textwidth}
		\includegraphics[width=\textwidth]{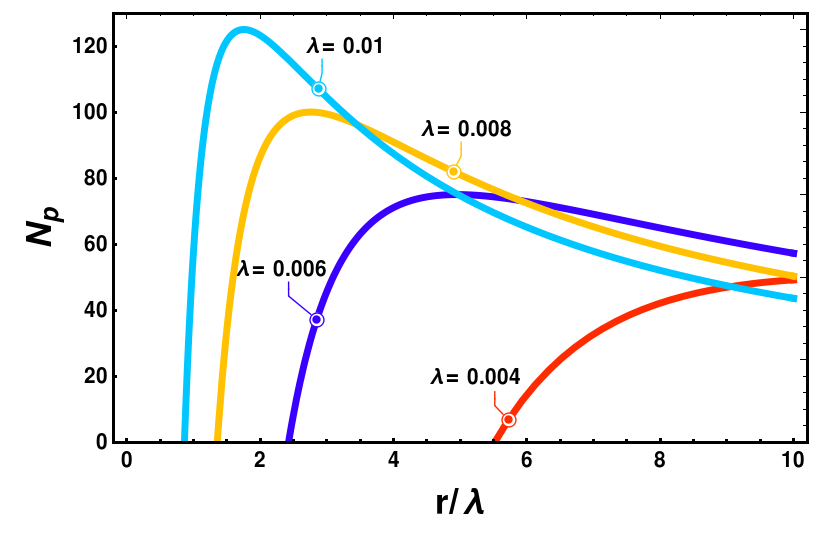}
		\caption{$\beta=1.1$}
		\label{varere2a}
	\end{subfigure}
	\hfill
	\begin{subfigure}{0.3\textwidth}
		\includegraphics[width=\textwidth]{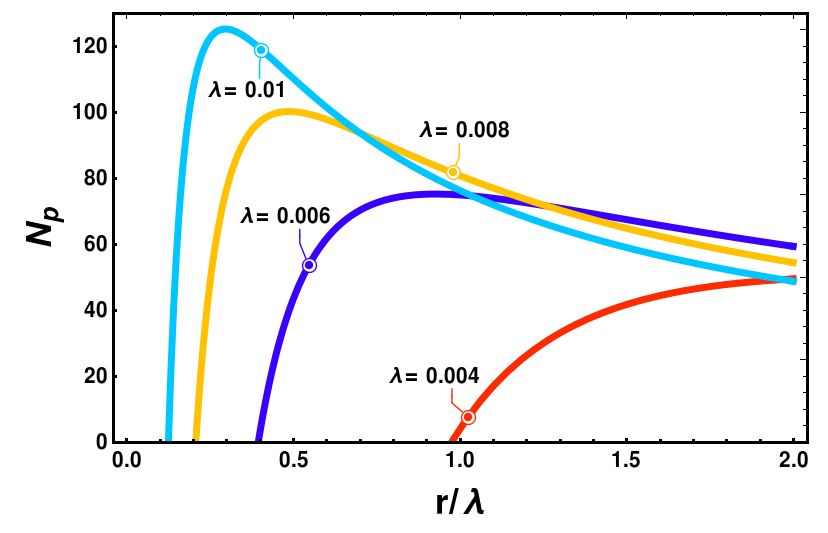}
		\caption{$\beta=2.0$}
		\label{varere2b}
	\end{subfigure}
	\hfill
	\begin{subfigure}{0.3\textwidth}
		\includegraphics[width=\textwidth]{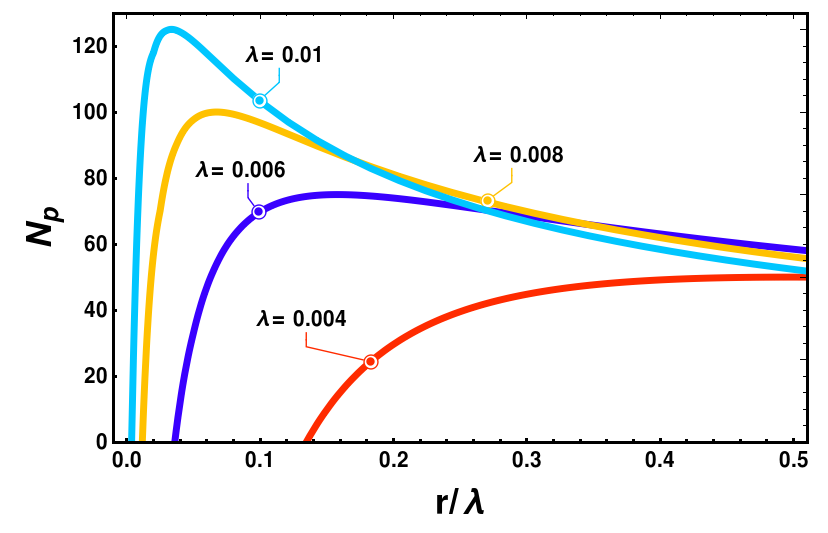}
		\caption{$\beta=3.0$}
		\label{varere2c}		
	\end{subfigure}
	\caption{Total population of predators as a function of the resetting rate, for several values of $\beta$ and low reproductive rates $\lambda$. (In all cases, $\mu=0.0001$).}
	\label{varere2}
\end{figure}

	In Figure \ref{variacionr}, $\mu$ is set to 0.01 (like in Fig. {\ref{fig:btaerea}}) whereas in Fig. {\ref{varere2}} it is set to 0.0001. These figures  display the total predator population as a function of $r$, for different choices of the L\'evy exponent $\beta$ and reproductive rate $\lambda$. For convenience, the re-scaled variable $r/\lambda$ is used in the figures.
    We observe a non-monotonic behaviour with a maximum at a particular $r_{opt}$, whose value depends on $\beta$ and $\lambda$, as expected from Eq. (\ref{findmaximum}). As $\lambda$ increases, $r_{opt}$ decreases (Fig \ref{varerea}): fast reproducing predators need to revisit less often the prey patch, otherwise they would over-exploit it. When $\beta$ is larger, namely, diffusion slower, fig. \ref{varereb}-c,  $r_{opt}$ tends to decrease, a property which can be interpreted by the fact that resetting is less needed because predators can often return to the origin by random movement. 
	\smallskip	
	
	\smallskip
    
\begin{figure}[htbp]
			\centering
			\begin{subfigure}{0.3\textwidth}
				\includegraphics[width=\textwidth]{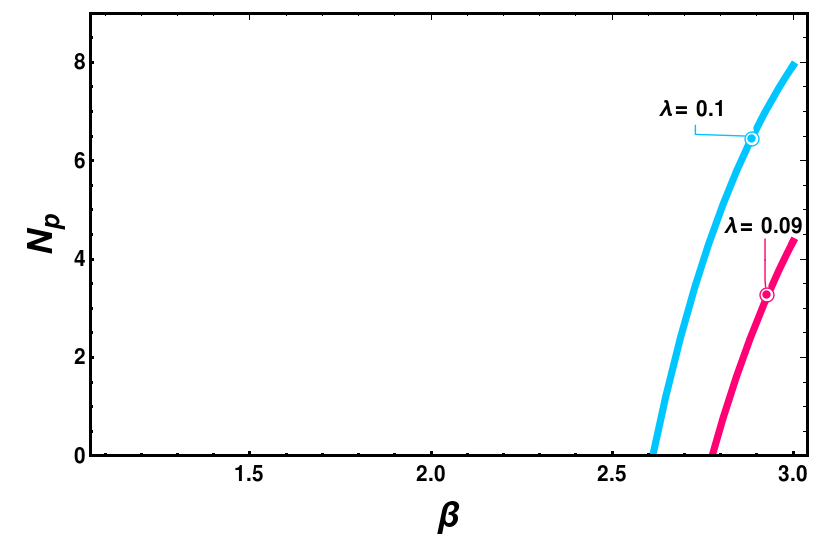}
				\caption{$r=0.0$}
				\label{varbta}
			\end{subfigure}
			\hfill
			\begin{subfigure}{0.3\textwidth}
				\includegraphics[width=\textwidth]{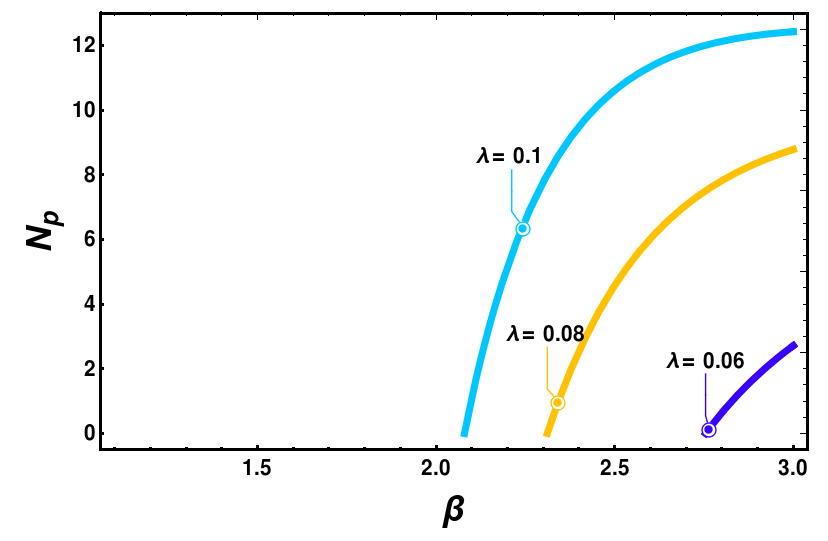}
				\caption{$r=0.01$}
				\label{varbtb}
			\end{subfigure}
			\hfill
			\begin{subfigure}{0.3\textwidth}
				\includegraphics[width=\textwidth]{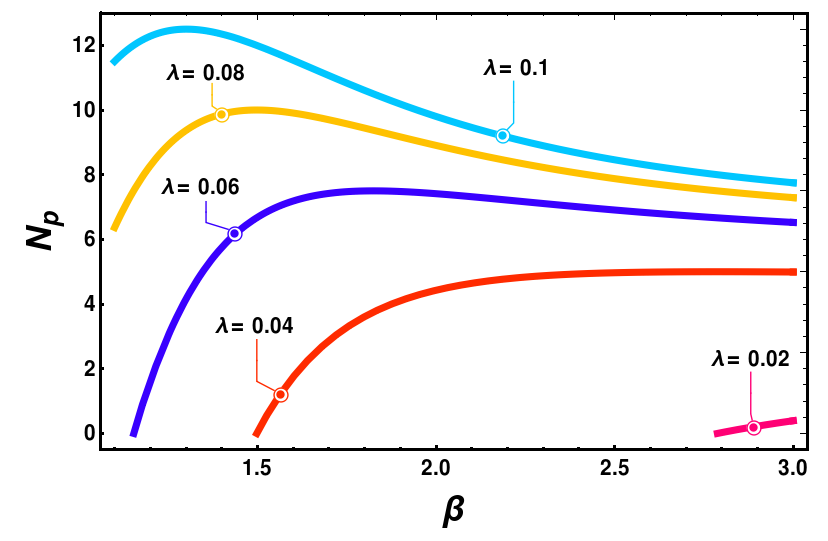}
				\caption{$r=0.15$}
				\label{varbtc}
			\end{subfigure}
			\caption{Total population of predators as a function of $\beta$, for several values of $r$ and reproductive rates $\lambda$. In all cases, $\mu=0.01$ (high mortality).}
			\label{varbt}
		\end{figure}

\begin{figure}[htbp]
			\centering
			\begin{subfigure}{0.3\textwidth}
				\includegraphics[width=\textwidth]{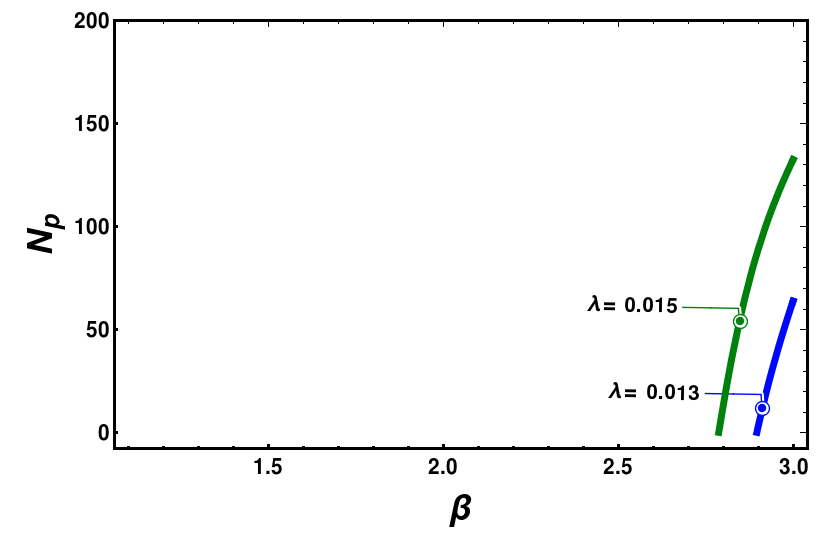}
				\caption{$r=0.0$}
				\label{varbt2a}
			\end{subfigure}
			\hfill
			\begin{subfigure}{0.3\textwidth}
				\includegraphics[width=\textwidth]{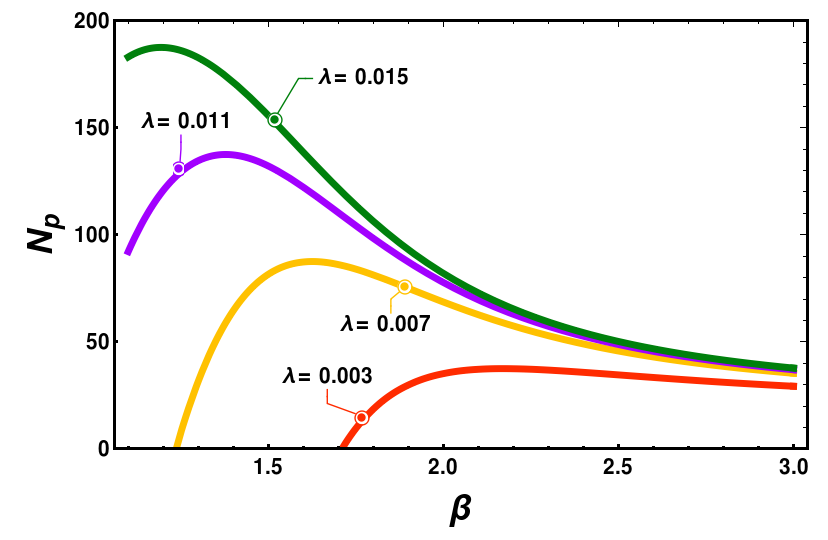}
				\caption{$r=0.01$}
				\label{varbt2b}
			\end{subfigure}
			\hfill
			\begin{subfigure}{0.3\textwidth}
				\includegraphics[width=\textwidth]{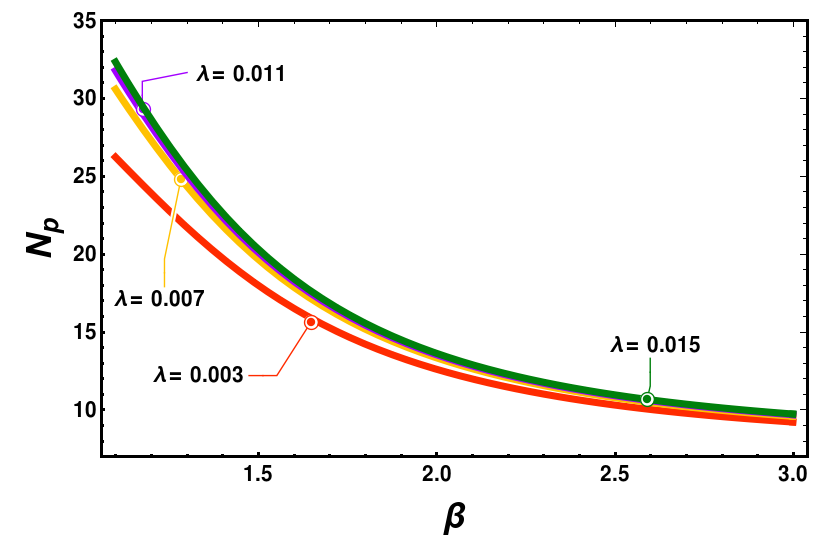}
				\caption{$r=0.15$}
				\label{varbt2c}
			\end{subfigure}
			\caption{Total population of predators as a function of $\beta$, for several values of $r$ and reproductive rates $\lambda$. In all cases, $\mu=0.0001$ (low mortality). }
			\label{varbt2}
		\end{figure}

    %% We can see the same behavior in the figures \ref{varere2a}-c as we explain before, thus we can think that there is a scale-free behavior of the position of $r^{*}$ respect to the reproduction rate.     
    Figures \ref{varbt} ($\mu=0.01$) and \ref{varbt2} ($\mu=0.0001$), display $N_p$ as a function of $\beta$, for different combinations of  $r$ and $\lambda$. The predator abundance is sometimes  non-monotonic in the L\'evy range and exhibits a maximum at a particular value $\beta_{opt}$. In general, the value of $\beta_{opt}$ maximizing the predator abundance can be in the L\'evy interval $(1,3)$, or be $>3$, or take the minimal value $1$. This exponent represents the optimal random movement strategy for the predator population as a whole. The three aforementioned regimes for the optimal value of $\beta$ were also observed in the model studied in \cite{pnas}, which considered $r=0$ and a finite patch density, {\it i.e.}, a finite separation distance between neighboring patches.
    
    In the absence of resetting (Fig. \ref{varbta}-\ref{varbt2a}), $\beta_{opt}$ is in the Brownian regime ($\beta_{opt}>3$) in the present single patch configuration. Too much super-diffusion causes predators not to return to the patch often enough, implying a population decline, and even extinctions at small $\beta$. When the reproduction rate $\lambda$ increases, so does $N_p$ and the range of $\beta$ allowing survival. In the many patch model of \cite{pnas}, $\beta_{opt}$ could be found in the L\'evy range for some intermediate values of the mortality and reproduction rates.
		\smallskip       
		
        When the resetting rate is switched on, the range of values of $\beta$ allowing survival is significantly wider, as predators always return to the prey patch soon or later to feed and reproduce. For this reason, at fixed $\lambda$ and for low values of $\beta$, increasing $r$ from $0$ has a positive effect on $N_p$ (Figs. \ref{varbtb}-c).  At fixed $\mu$, the predation pressure can be increased by making $\lambda$ larger and the diffusion slower ({\it i.e.} $\beta$ larger): in this case, increasing $r$ implies a decrease of $N_p$ (compare the top curve of Figs. \ref{varbt} b and c for $\beta\sim 3$). 
 At low mortality, this effect is more drastic, as can be noticed by the difference in the vertical scales between Fig. \ref{varbt2a} ($r=0$) and  Fig. \ref{varbt2c} ($r=0.15$).
 Meanwhile, $\beta_{opt}$ decreases when $\lambda$ increases: the over-exploitation of resources is avoided by an optimal movement mode that becomes more super-diffusive. We thus conclude that resetting has a greater impact on the survival of highly super-diffusive L\'evy populations than on Brownian populations.
        
        A comparison of Figs. \ref{varbtb}-c and \ref{varbt2b}-c also reveals that $\beta_{opt}$ decreases as $r$ increases. The bell-shape can even disappear and be replaced by a monotonic decaying from $\beta_{opt}$=1 instead. 
		\smallskip
        
    \subsection{Parameters maximizing predator populations}

It follows from the above results that the predator abundance reaches a maximum when the movement parameter $\beta$ is set to some value $\beta_{opt}$, which can be in the L\'evy range  $1<\beta_{opt}<3$. We proceed below to better locate the region of optimal L\'evy flights in parameter space.

\begin{figure}[htbp]
		\centering
        
        \begin{subfigure}{0.3\textwidth}
				\includegraphics[width=\textwidth]{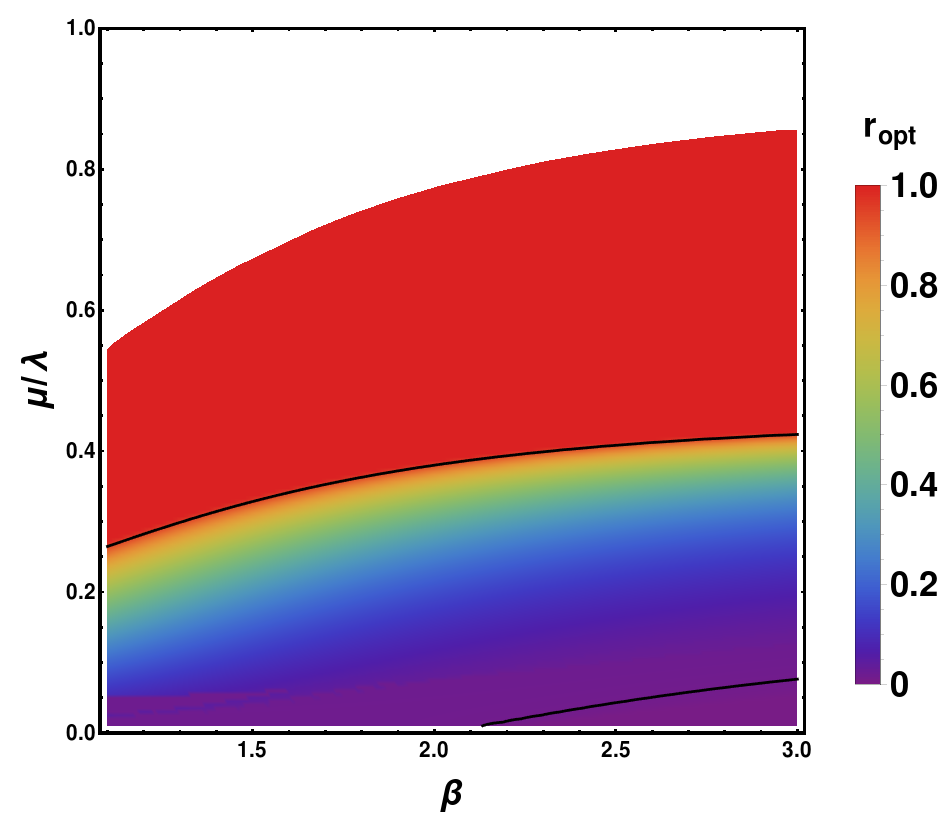}
				\caption{$\lambda=0.1$}
				\label{maxerea}
			\end{subfigure}
%			\begin{subfigure}{0.3\textwidth}
%				\includegraphics[width=\textwidth]{rmaxdensitynormal3}
%				\caption{$\lambda=0.3$}
%				\label{maxereb}
%			\end{subfigure}
		\begin{subfigure}{0.3\textwidth}
			\includegraphics[width=\textwidth]{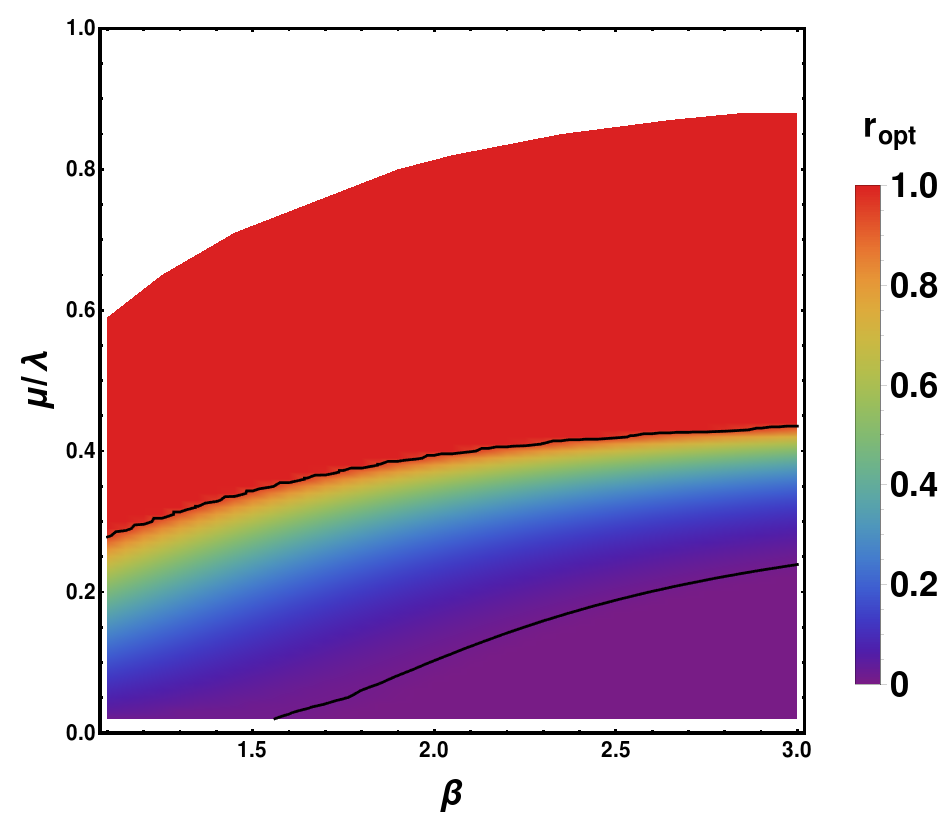}
			\caption{$\lambda=0.5$}
			\label{maxerec}
		\end{subfigure}
        
		\begin{subfigure}{0.3\textwidth}
			\includegraphics[width=\textwidth]{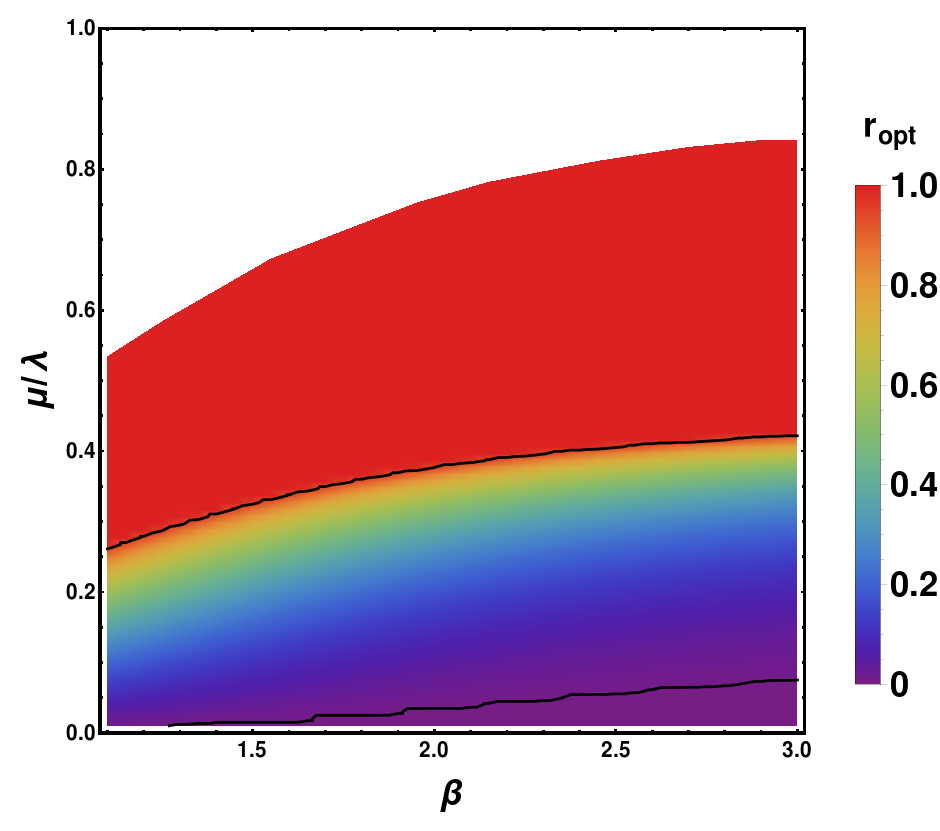}
			\caption{$\lambda=0.0001$}
			\label{maxere2a}
		\end{subfigure}
%		\begin{subfigure}{0.3\textwidth}
%			\includegraphics[width=\textwidth]{rmaxdensitynormal0003}
%			\caption{$\lambda=0.0003$}
%			\label{maxere2b}
%		\end{subfigure}
		\begin{subfigure}{0.3\textwidth}
			\includegraphics[width=\textwidth]{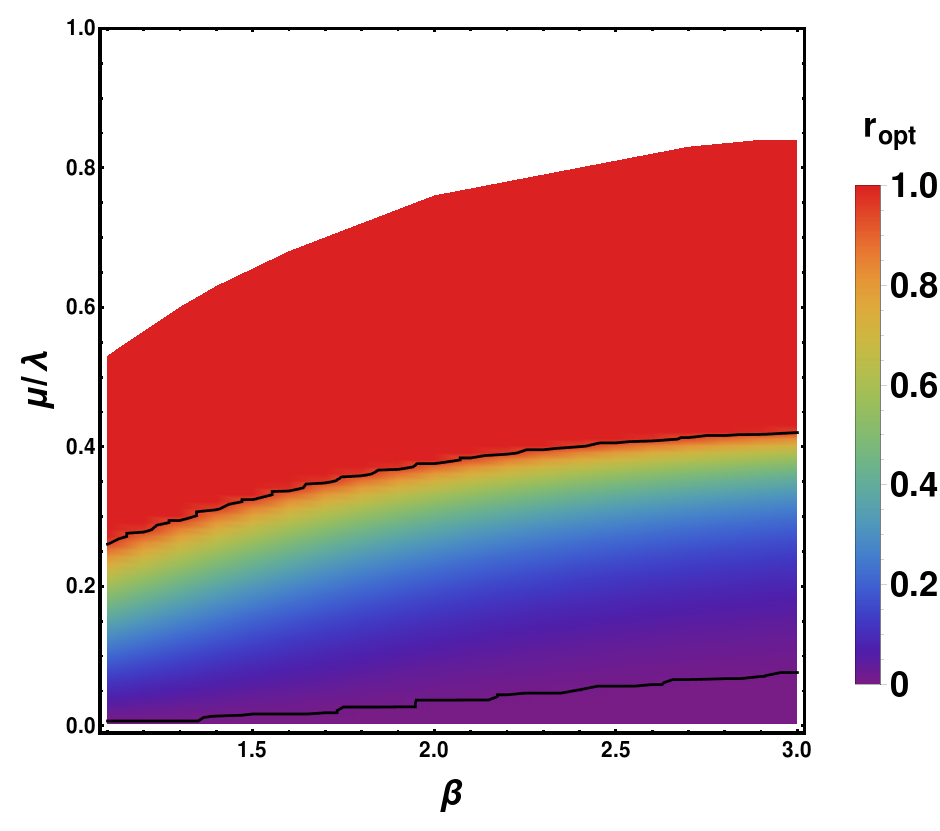}
			\caption{$\lambda=0.0005$}
			\label{maxere2c}
		\end{subfigure}
		\caption{Stochastic resetting parameter $r$ that maximizes $N_p$ given $\beta$ and $\mu$, for $4$ different values of $\lambda$. Above the upper solid line (red region), $r_{opt}>1$, and below the lower solid line, $r_{opt}<0.01$. The white region corresponds to predator extinction or $N_p=0$.}
		\label{maxere2}
	\end{figure}

\begin{figure}[htbp]
			\centering
            
            \begin{subfigure}{0.3\textwidth}
				\includegraphics[width=\textwidth]{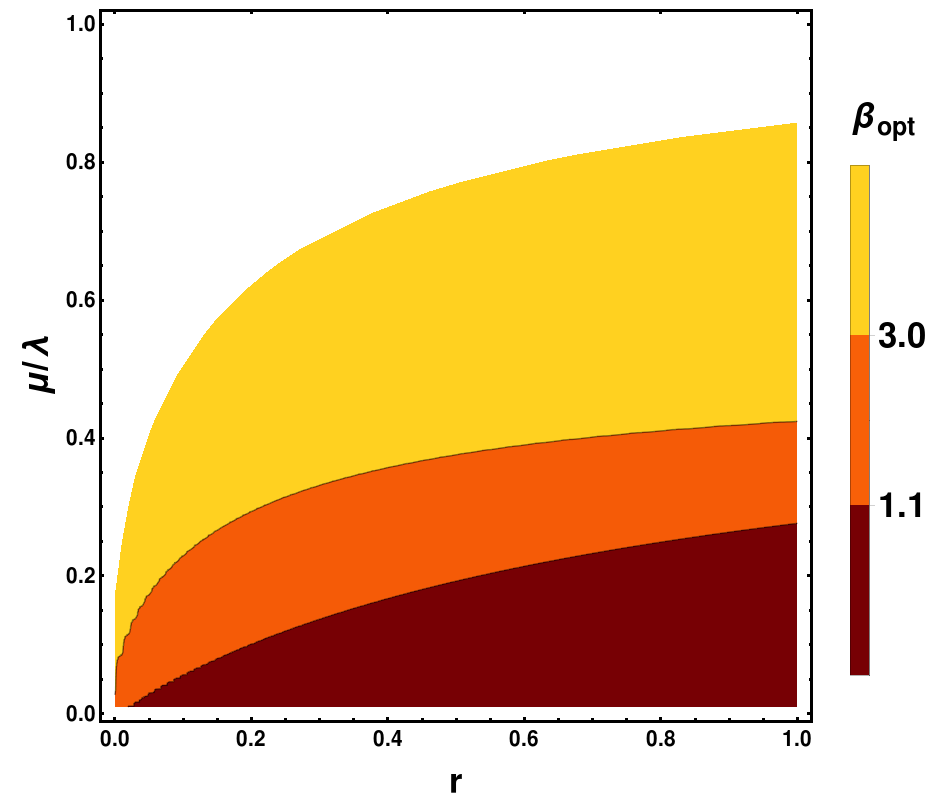}
				\caption{$\lambda=0.1$}
				\label{maxbtaa}
			\end{subfigure}
			\begin{subfigure}{0.3\textwidth}
				\includegraphics[width=\textwidth]{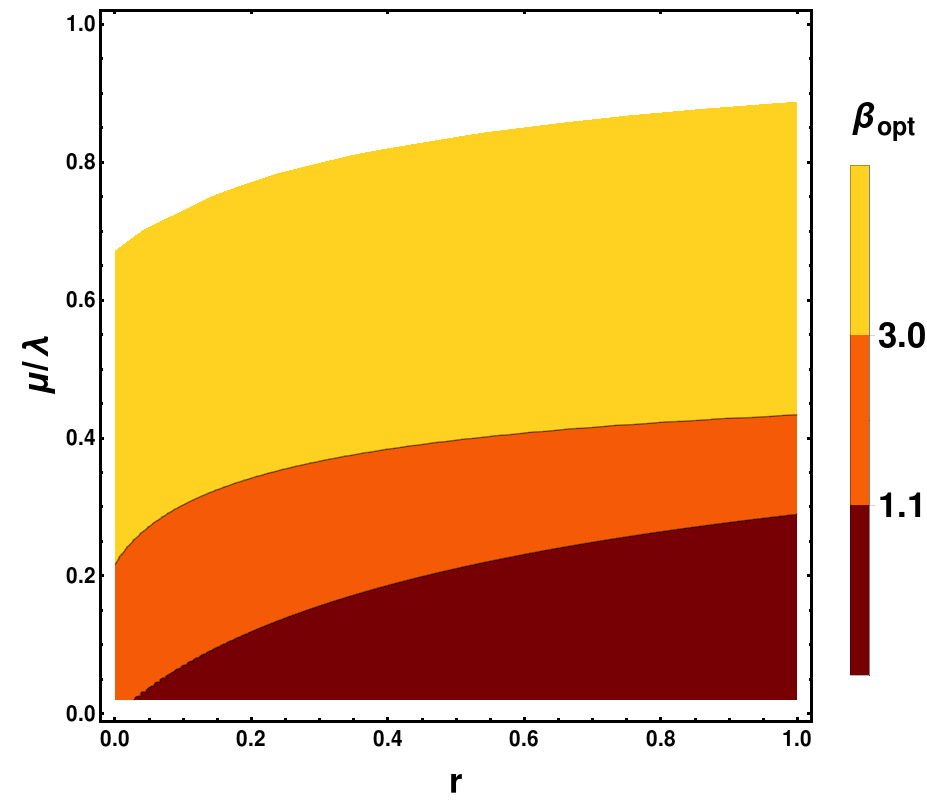}
				\caption{$\lambda=0.5$}
				\label{maxbtac}
			\end{subfigure}
            
			\begin{subfigure}{0.3\textwidth}
				\includegraphics[width=\textwidth]{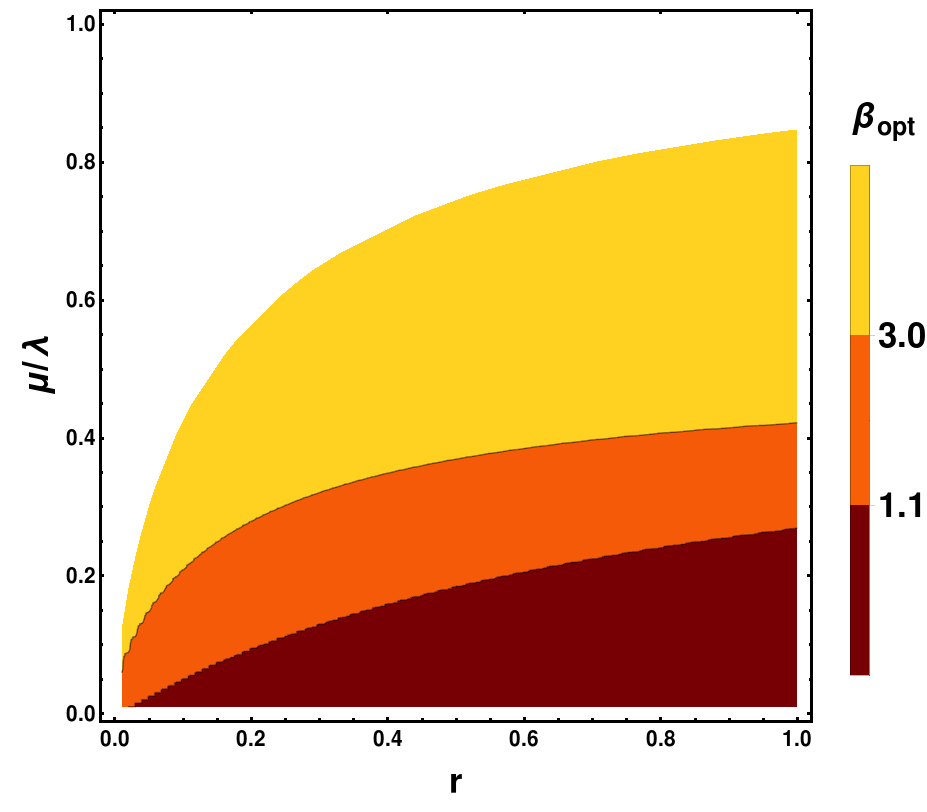}
				\caption{$\lambda=0.0001$}
				\label{maxbta2a}
			\end{subfigure}
			\begin{subfigure}{0.3\textwidth}
				\includegraphics[width=\textwidth]{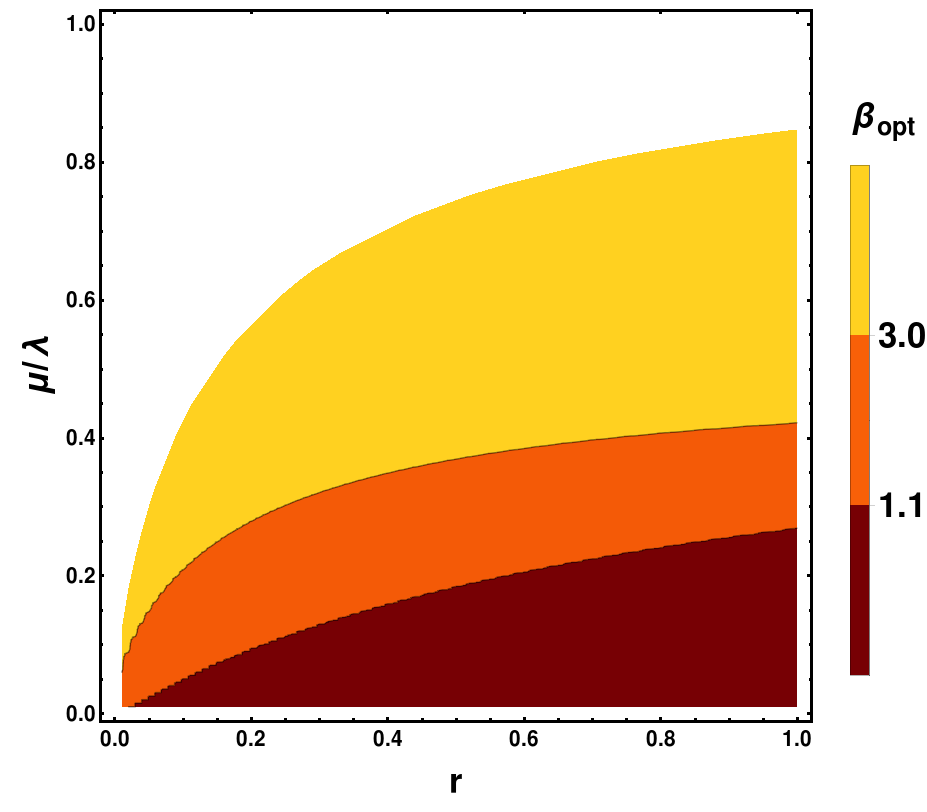}
				\caption{$\lambda=0.0005$}
				\label{maxbta2c}
			\end{subfigure}
			\caption{The parameter $\beta$ that maximizes $N_p$ given $r$ and $\mu$ is displayed. We set different values of $\lambda$. The white region corresponds to predator extinction or $N_p=0$.}
			\label{maxbta2}
		\end{figure}

		We wish to find the movement parameters $\beta$ and $r$ that maximize the forager abundance at given $\mu$ and $\lambda$. Due to the difficulty of representing this domain in a three-dimensional frame, we fix $\lambda$ and vary the other three parameters in two different ways. In the first case, we fix $\mu$ and $\beta$ in the L\'evy range $(1,3)$, find the $r_{opt}$ that maximizes $N_p$ and repeat the operation for many pairs $(\mu,\beta)$, as shown in Figure \ref{maxere2}. In the second case, we fix $(\mu,r)$ and determine $\beta_{opt}$, as shown in Fig. \ref{maxbta2}. In the later case $\beta_{opt}$ is not necessarily in the L\'evy range, thus we can address the question whether L\'evy strategies are often optimal or requires some particular conditions.

Figure \ref{maxere2} exhibits in color-scale the optimal $r_{opt}$ as a function of $\mu$ and $\beta$ ($\lambda$ being set to 4 different values). When using the re-scaled variable $\mu/\lambda$, all graphs are very similar. 
As previously noticed, non-zero values of $r$ are advantageous for super-diffusive predators in the whole range of $\mu$. When the mortality is very high ($\mu>\lambda/4$) the maximum population is even reached with $r^*\gtrsim1$ irrespective to the value of $\beta$, meaning that the foragers reset so often that they practically remain localized on the prey patch. Long lived $(\mu\ll\lambda)$ and not-so-super-diffusive foragers $(\beta>2)$ require a small resetting rate to maximize their population.
		\smallskip
		
	In Figure \ref{maxbta2}, the optimal $\beta$ is classified into the three regimes $\beta_{opt}>3$ (Brownian), L\'evy, and $1<\beta_{opt}<1.1$ (highly-superdiffusive). Brownian strategies are advantageous for values of $\mu/\lambda$ above a certain threshold and below the extinction threshold ($N_p=0$ in the white region). When $r$ is small, the former threshold is  fairly small, see Fig. \ref{maxbta2}a,c,d. At lower mortality rate, L\'evy flights become advantageous since they avoid overpopulation on the patch. Interestingly, L\'evy strategies become optimal when both $r$ and $\mu/\lambda$ become very small. The highly super-diffusive regime is never favored at low $r$, but become vastly optimal at larger $r$ and small $\mu/\lambda$ (see Fig. \ref{varbt2c} for an example).
		\bigskip
		
\subsection{Stochastic 2D simulations}

\begin{figure}[htbp]
			\centering
        	\begin{subfigure}{0.4\textwidth}
			\includegraphics[width=\textwidth]{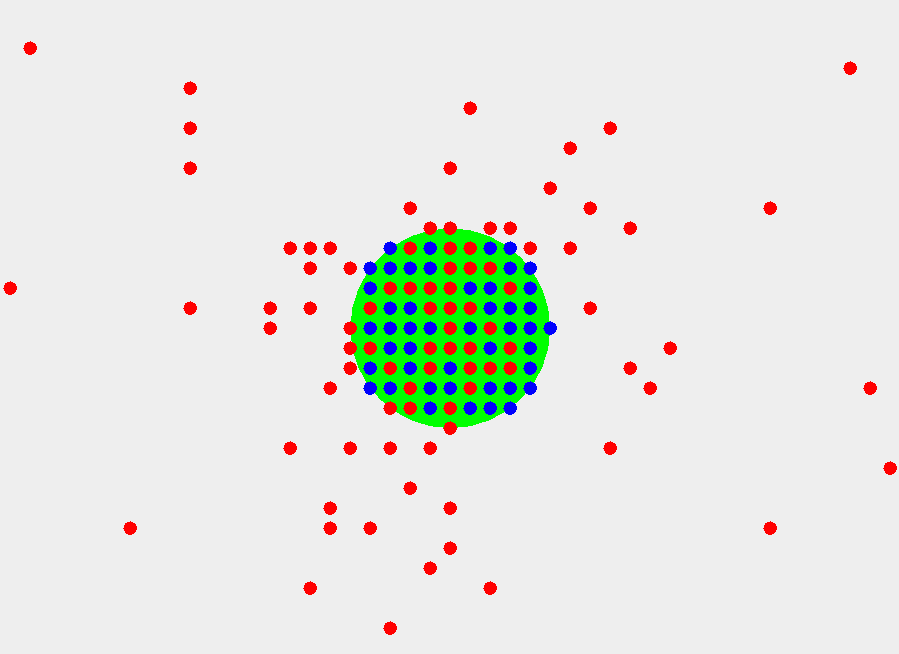}
			\caption{$\lambda=0.1$, $\mu=0.01$, $r=0.05$, $\beta=2$}
			\label{simuProgram}
		\end{subfigure}
        \hspace{1cm}
        \begin{subfigure}{0.4\textwidth}
			\includegraphics[width=\textwidth]{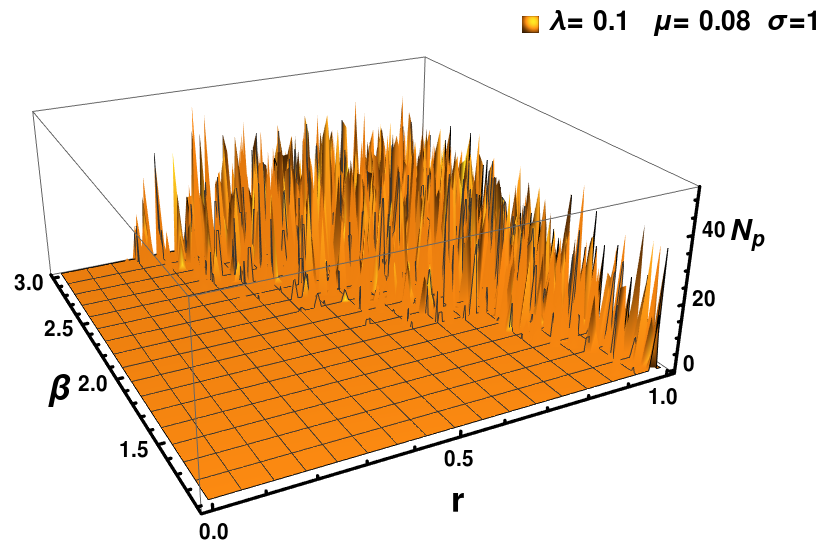}
			\caption{$\lambda=0.1$, $\mu=0.08$ and $\sigma=1$}
			\label{simuGraph8}
		\end{subfigure}
        \hspace{1cm}
        \begin{subfigure}{0.4\textwidth}
			\includegraphics[width=\textwidth]{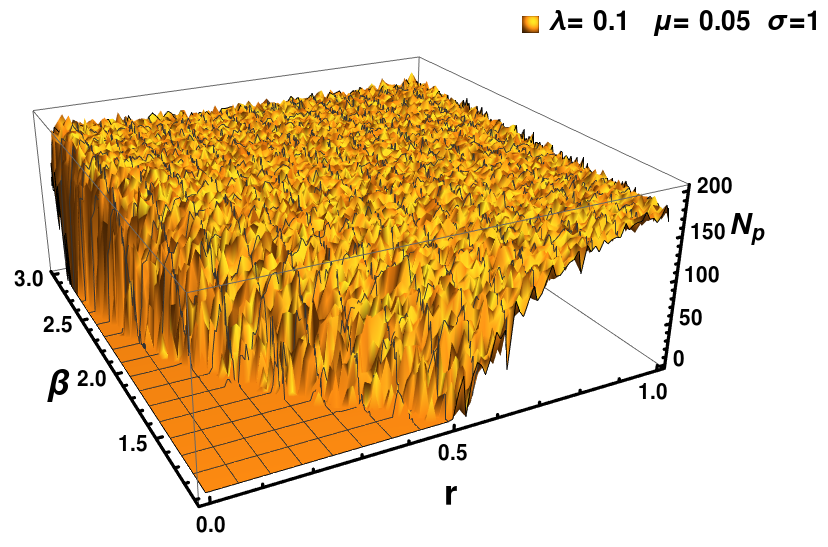}
			\caption{$\lambda=0.1$, $\mu=0.05$ and $\sigma=1$}
			\label{simuGraph5}
		\end{subfigure}
        \hspace{1cm}
        \begin{subfigure}{0.4\textwidth}
			\includegraphics[width=\textwidth]{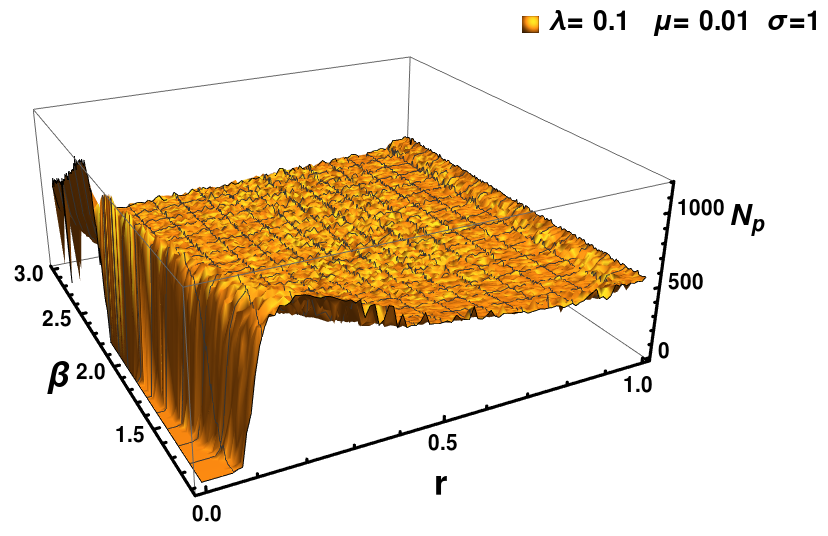}
			\caption{$\lambda=0.1$, $\mu=0.01$ and $\sigma=1$}
			\label{simuGraph1}
		\end{subfigure}
		\caption{a) In simulations, space is a two-dimensional discretized square lattice, with some sites forming a patch in the center (green area). Inside the patch, a site can contain at most one prey (blue dots). There is no limitation on the number of predators (red dots) per site. Predators perform independent discrete 2D L\'evy flights. When a predator (red dots) arrives at a site containing a prey, it consumes it with probability $\lambda'=1$ and reproduces with probability $\lambda$.  In one Monte Carlo time step, predators die with probability $\mu$ whereas surviving predators perform one movement step.  The patch radius is set to $R=10$. An empty site in the patch next to a prey can become occupied by a new prey with probability $\sigma$. The curves display the predator abundance as a function of the movement parameters $\beta$ and $r$ with $\mu=0.08$ (b), $\mu=0.05$ (c) and $\mu=0.01$ (d), for a fixed $\lambda=0.1$. }
		\label{simulation}
	\end{figure}

To illustrate the usefulness of the one-dimensional cases considered so far to infer properties of two-dimensional systems, we have performed Monte Carlo simulations of a 2D stochastic population model that is the stochastic analogue of Eqs. (\ref{lotka})-(\ref{preydensity}) for the species densities. At high densities, the stochastic and deterministic descriptions become equivalent \cite{Mobilia}, but when the number of individuals is small, fluctuations can play an important role. We have implemented an algorithm (see caption of Figure {\ref{simuProgram}}) which is also exposed with more details in Refs. \cite{Mobilia} and \cite{pnas}. 

The curves of predator abundance in 2D, displayed in Figs. {\ref{simuGraph8}}-c-d, are similar to the ones shown in Figs. {\ref{btaere}}a-b-c for the 1D case and the same parameter values. The only difference is the value of $R$, which is the length of a cell in 1D and the radius of a circular patch in 2D. In Figure {\ref{simuGraph8}}, where $\mu/\lambda\sim 1$, the total number of predators is small and subject to large fluctuations due to stochasticity. At smaller $\mu/\lambda$ (Fig. {\ref{simuGraph5}}), the monotonic behaviour with $r$ and $\beta$ becomes noticeable, whereas $N_p$ remains $0$ at small $\beta$ and $r$. These results are qualitatively similar to those of the deterministic model (Fig. {\ref{btaere}}a-b). In Fig. {\ref{simuGraph1}}, where $\mu/\lambda=0.1$, one recovers as well the most salient features of the analytical model in this regime (Fig. {\ref{btaere}}c): namely, the non-monotonic behaviour with $r$, the decrease of $r_{opt}$ with $\beta$ and the persistence of extinct states at small $\beta$ and $r$.

	\section{Discussion and conclusions}
 In this work, we have analysed the properties of diffusing elements with stochastic resetting to the origin, that interact through a population dynamics model. We have considered more specifically a predator/prey system where prey are scarce, that is to say, confined to the origin cell (patch). Predators reproduce on the prey patch and randomly migrate from cell to cell on the lattice according to a power-law dispersal kernel with exponent $\beta$. The fact that predators stochastically relocate (reset) from time to time to the origin is motivated by site fidelity, an important aspect of foraging ecology \cite{spencer1990operationally,foerster2002home,weimerskirch2007seabirds,wood2017home}. Our model admits asymptotic stationary distributions for the species densities and we have obtained analytical expressions for the species abundances.
    
    \smallskip
    
    Fixing the demographic parameters, chiefly, the predator mortality and reproduction rates, there exists in general many pairs of movement parameters ($\beta$ and the resetting rate $r$) for which the predator population is maximal. Such foraging strategies thus allows an optimal exploitation of resources at the population level. This collective notion of optimality (also discussed in \cite{pnas}) differs from the more usual individual optimality that is studied in search models, which is based on maximising the rate of prey capture by a single agent in an environment of static prey. In these contexts, interactions between agents and birth/death processes are usually ignored. 

\smallskip

The predator abundance represents a natural quantity to measure the foraging efficiency from a collective point of view. In the deterministic approach that we use, this quantity evolves toward a stable steady state solution (there are no oscillations in our model). In a stochastic approach, however, there are other ways of quantifying the efficiency, such as the mean first passage time to extinction or to a threshold value of prey population which allows the population of predators to survive. We did not focus on stochasticity in this work, although it is an important aspect deserving future research.

 \smallskip
 
A first category of unfavorable environmental conditions corresponds to predators with high mortality and slow reproduction. In this case, as expected, the foraging strategy consisting of diffusing normally and with a high resetting rate to the prey patch allows a better exploitation of resources and generates more abundant forager populations (Figs. \ref{fig:btaerec}-\ref{fig:btaereb}). This situation changes when the predator mortality is too low: in this second category, the risk is to over-populate the prey cell and under-populate the other cells. In such cases, a low resetting rate is preferable, since it gives the time for foragers to diffuse away, thus increasing the prey density on the patch and consequently the reproduction rate per predator. Comparatively, the mobility exponent $\beta$ plays a less important role: L\'evy and Brownian foragers can reach the same maximal population levels, owing to the logistic dependence of $N_p$ on $a_0$, although at different optimum resetting rates
(Fig. \ref{fig:btaerea}).  In Fig. \ref{variacionr}-\ref{varere2}, one can appreciate that the abundances at their maximum as a function of $r$ do not depend on $\beta$. At even smaller mortality rate (compared to the reproduction rate), the optimal resetting rate tends to zero for any value of $\beta$ (Figs. \ref{variacionr}-\ref{varere2}).

\smallskip

The total predator abundance is more sensitive to the L\'evy exponent $\beta$ when $r$ is fixed to some value, which is not necessarily the optimal one (Figs. \ref{varbt}-\ref{varbt2}). Generally speaking, at fixed resetting rate, the optimal $\beta$ moves from 1 (highly super-diffusive) to a value $>3$ (Brownian) as the relative mortality rate $\mu/\lambda$ increases. Hence, at small mortality, the over-population of the origin is avoided thanks to super-diffusion between resetting events. It is important to note that, when $r=0$, one would recover the model studied in\cite{pnas} in the limit of vanishing patch density. However, in this previous work, only finite patch densities were considered, hence the results of \cite{pnas}, cannot be directly compared with the currents ones, in particular Figs. {\ref{varbt}}a and {\ref{varbt2}}a. However, the non-monotonic behaviour of the predator abundance with $\beta$, when $r$ is different from zero, is similar to the behaviour found in Fig. 5 of\cite{pnas}, where $\beta_{opt}$ could also lie in the L\'evy range. It is possible that, in a first approximation, the average distance travelled before resetting in the current model is analogous to the average between-patch distance in \cite{pnas}.

\smallskip
 
In the particular case corresponding to the absence of resetting ($r=0$), predators will avoid extinction if $\beta>\beta_c$, {\it i.e.}, if they are not excessively super-diffusive (Figs. \ref{varbta}-\ref{varbt2a}). Otherwise, the probability of returning to the origin by chance during a lifetime ($1/\mu$) is too small and this effect is not compensated by the birth of new individuals within the patch. One of our main conclusion is that setting $r$ different from zero greatly improves the abundance and widens the range of values of $\beta$ allowing survival in one-patch configurations. 

\smallskip

Conversely, if $\beta$ is held fixed in the L\'evy regime, there is an critical resetting rate $r_c$ below which no population survive (Figs. \ref{variacionr}-\ref{varere2}). This critical value decreases as foragers become more Brownian, since they become more recurrent and need less resetting to occupy the origin and reproduce. In addition, at small enough $\mu/\lambda$, $N_p$ is non-monotonous: there exist an optimal resetting rate that maximizes the predator abundance. This finding is reminiscent of the non-monotonic behaviour with respect to $r$ of the first passage time to a target site for a diffusing particle initially located at a given distance \cite{PhysRevLett.106.160601,evans2011diffusion,camposmendez2015,pal2016}. In our context, too much resetting produces excessive concentration on a single patch and resource over-exploitation, whereas too few resetting generates few predation and reproduction. 
 
 \smallskip
 
It is satisfying to notice that the conclusions obtained for 1D systems depend little on the spatial dimension, as shown by the 2D stochastic simulation results. We thus expect the 1D cases to give useful insights on patch configurations that are ecologically more realistic. Future extensions of the present work should consider several prey patches randomly distributed in space, as studied in \cite{pnas}, and where resetting would take place to different places. The exponent $\beta$ could play a significant role in multiple patch systems, since super-diffusion may allow the colonization of distant patches. In future work, predators could perform L\'evy walks instead of L\'evy flights, since this former model has been widely used in the context of animal movement and random search problems  \cite{viswanathan1999optimizing,gandhibook,bartulevin}. 
 
 \bigskip
 
%Dentro de las extensiones al modelo que se proponen sugerimos establecer distintos puntos de retorno en la caminata del depredador, de manera que el reseteo estoc\'astico no est\'e condicionado al origen sino podr\'ia ser a distintos parches, intentando as\'i describir el movimiento en busca de alimento que se caracteriza en el trap-lining, de manera que se pueda establecer una estrategia \'optima para la supervivencia de una especie que a localizado fuentes de alimento en una determinada zona.
  
 %Sistemas ecol\'ogicos en los cuales el depredador vive en una determinada area de donde obtiene todos los recursos necesarios para su supervivencia han sido tratados utilizando diferentes enfoques, mediante caminatas aleatorias no-Markovianas o mediante modelos de difusi\'on. En este contexto, nuestro trabajo ofrece una posibilidad de describir este tipo de depredadores hogare\~nos que realizan una busqueda de alimento dentro de una regi\'on delimitada, saliendo de esta zona durante su busqueda para invariablemente retornar a un lugar de origen.
   %%\cite{PhysRevLett.112.240601,van2009memory,borger2008there,moorcroft2006mechanistic}%%

{\bf Acknowledgements}

\smallskip

We acknowledge support from PAPIIT Grants IN105015 (DB) and IN108318 (DB and JGMV) of the Universidad Nacional Aut\'onoma de M\'exico. JGMV also thanks CONACYT (M\'exico) for a Masters scholarship. 

%\bibliographystyle{IEEEtran}
%\bibliographystyle{iopart-num}
%\bibliography{regeneracion}

\begin{thebibliography}{10}
\expandafter\ifx\csname url\endcsname\relax
  \def\url#1{{\tt #1}}\fi
\expandafter\ifx\csname urlprefix\endcsname\relax\def\urlprefix{URL }\fi
\providecommand{\eprint}[2][]{\url{#2}}
% Bibliography created with iopart-num v2.1
% /biblio/bibtex/contrib/iopart-num

\bibitem{Ilkka1998}
Hanski I 1998 {\em Nature\/} {\bf 396} 41--49

\bibitem{Turchin98}
Turchin P 1998 {\em Quantitative Analysis of Movement: Measuring and Modeling
  Population Redistribution in Animals and Plants\/} 1st ed (Sinauer Associates
  Inc)

\bibitem{nathan2008movement}
Nathan R, Getz W~M, Revilla E, Holyoak M, Kadmon R, Saltz D and Smouse P~E 2008
  {\em Proceedings of the National Academy of Sciences\/} {\bf 105}
  19052--19059

\bibitem{metapopulationDiekmann}
Sabelis M~W, Diekmann O and Jansen V~A~A 1991 {\em Biological Journal of the
  Linnean Society\/} {\bf 42} 267--283

\bibitem{holland2008strong}
Holland M~D and Hastings A 2008 {\em Nature\/} {\bf 456} 792

\bibitem{huffaker1958experimental}
Huffaker C {\em et~al.\/} 1958 {\em California Agriculture\/} {\bf 27} 343--383

\bibitem{mcmurtrie1978persistence}
McMurtrie R 1978 {\em Mathematical Biosciences\/} {\bf 39} 11--51

\bibitem{populationbiologyhastings}
Hastings A 1997 {\em Population Biology: Concepts and Models\/} 1st ed
  (Springer-Verlag New York)

\bibitem{maynard1978models}
Maynard-Smith J 1978 {\em Models in ecology\/} (CUP Archive)

\bibitem{solebascompte}
Bascompte J and Sol{\'e} R~V 1996 {\em Journal of Animal Ecology\/}  465--473

\bibitem{Mobilia}
Mobilia M, Georgiev I~T and Tauber U~C 2007 {\em J. Stat. Phys.\/} {\bf 128}

\bibitem{bartumeus2016foraging}
Bartumeus F, Campos D, Ryu W~S, Lloret-Cabot R, M{\'e}ndez V and Catalan J 2016
  {\em Ecology Letters\/} {\bf 19} 1299--1313

\bibitem{switzer1993site}
Switzer P~V 1993 {\em Evolutionary Ecology\/} {\bf 7} 533--555

\bibitem{gandhi1999}
Viswanathan G~M, Buldyrev S~V, Havlin S, Da~Luz M, Raposo E and Stanley H~E
  1999 {\em Nature\/} {\bf 401} 911--914

\bibitem{gandhibook}
Viswanathan G~M, Da~Luz M~G, Raposo E~P and Stanley H~E 2011 {\em The physics
  of foraging: an introduction to random searches and biological encounters\/}
  (Cambridge University Press)

\bibitem{bartu2003}
Bartumeus F, Peters F, Pueyo S, Marras{\'e} C and Catalan J 2003 {\em
  Proceedings of the National Academy of Sciences\/} {\bf 100} 12771--12775

\bibitem{RF}
Ramos-Fern{\'a}ndez G, Mateos J, Miramontes O, Cocho G, Larralde H and
  Ayala-Orozco B 2004 {\em Behavioral Ecology and Sociobiology\/} {\bf 55}
  223--230

\bibitem{boyermonos}
Boyer D, Ramos-Fern{\'a}ndez G, Miramontes O, Mateos J~L, Cocho G, Larralde H,
  Ramos H and Rojas F 2006 {\em Proceedings of the Royal Society of London B:
  Biological Sciences\/} {\bf 273} 1743--1750

\bibitem{reynolds}
Reynolds A~M, Smith A~D, Menzel R, Greggers U, Reynolds D~R and Riley J~R 2007
  {\em Ecology\/} {\bf 88} 1955--1961

\bibitem{Brown}
Brown C~T, Liebovitch L~S and Glendon R 2007 {\em Human Ecology\/} {\bf 35}
  129--138

\bibitem{Sims}
Sims D~W, Southall E~J, Humphries N~E, Hays G~C, Bradshaw C~J, Pitchford J~W,
  James A, Ahmed M~Z, Brierley A~S, Hindell M~A {\em et~al.\/} 2008 {\em
  Nature\/} {\bf 451} 1098--1102

\bibitem{deJager}
de~Jager M, Weissing F~J, Herman P~M, Nolet B~A and van~de Koppel J 2011 {\em
  Science\/} {\bf 332} 1551--1553

\bibitem{humphries}
Humphries N~E, Weimerskirch H, Queiroz N, Southall E~J and Sims D~W 2012 {\em
  Proceedings of the National Academy of Sciences\/} {\bf 109} 7169--7174

\bibitem{chechkin2008introduction}
Chechkin A~V, Metzler R, Klafter J, Gonchar V~Y {\em et~al.\/} 2008 {\em
  Anomalous transport: Foundations and Applications\/}  129--162

\bibitem{Jespersen2000}
Jespersen S and Blumen A 2000 {\em Phys. Rev. E\/} {\bf 62}(5) 6270--6274
  \urlprefix\url{https://link.aps.org/doi/10.1103/PhysRevE.62.6270}

\bibitem{Kozma2007}
Kozma B, Hastings M~B and Korniss G 2007 {\em Journal of Statistical Mechanics:
  Theory and Experiment\/} {\bf 2007} P08014
  \urlprefix\url{http://stacks.iop.org/1742-5468/2007/i=08/a=P08014}

\bibitem{Szabo2004}
Szab\'o G, Szolnoki A and Izs\'ak R 2004 {\em Journal of Physics A:
  Mathematical and General\/} {\bf 37} 2599
  \urlprefix\url{http://stacks.iop.org/0305-4470/37/i=7/a=006}

\bibitem{Shabunin2008}
Shabunin A and Efimov A 2008 {\em The European Physical Journal B\/} {\bf 65}
  387--393 ISSN 1434-6036
  \urlprefix\url{https://doi.org/10.1140/epjb/e2008-00316-5}

\bibitem{kerr2002local}
Kerr B, Riley M~A, Feldman M~W and Bohannan B~J 2002 {\em Nature\/} {\bf 418}
  171

\bibitem{reichenbach2007mobility}
Reichenbach T, Mobilia M and Frey E 2007 {\em Nature\/} {\bf 448} 1046

\bibitem{fagan2013spatial}
Fagan W~F, Lewis M~A, Auger-M{\'e}th{\'e} M, Avgar T, Benhamou S, Breed G,
  LaDage L, Schl{\"a}gel U~E, Tang W~w, Papastamatiou Y~P {\em et~al.\/} 2013
  {\em Ecology Letters\/} {\bf 16} 1316--1329

\bibitem{thomson1997trapline}
Thomson J~D, Slatkin M and Thomson B~A 1997 {\em Behavioral Ecology\/} {\bf 8}
  199--210

\bibitem{EcologyThomas1982}
Thomas H Kunz~(auth) T~H~K~e 1982 {\em Ecology of Bats\/} 1st ed (Springer US)

\bibitem{champanAnimal1990}
Bell W~J 1990 {\em Searching Behaviour: The behavioural ecology of finding
  resources\/} 1st ed Chapman and Hall Animal Behaviour Series (Springer
  Netherlands)

\bibitem{PhysRevLett.106.160601}
Evans M~R and Majumdar S~N 2011 {\em Physical Review Letters\/} {\bf 106}(16)
  160601

\bibitem{evans2011diffusion}
Evans M~R and Majumdar S~N 2011 {\em Journal of Physics A: Mathematical and
  Theoretical\/} {\bf 44} 435001

\bibitem{mendezcampos2016}
M\'endez V~m~c and Campos D 2016 {\em Phys. Rev. E\/} {\bf 93}(2) 022106

\bibitem{pal2016}
Pal A, Kundu A and Evans M~R 2016 {\em Journal of Physics A: Mathematical and
  Theoretical\/} {\bf 49} 225001

\bibitem{kusmierz2014first}
Kusmierz L, Majumdar S~N, Sabhapandit S and Schehr G 2014 {\em Physical Review
  Letters\/} {\bf 113} 220602

\bibitem{kusmierz2015optimal}
Ku{\'s}mierz {\L} and Gudowska-Nowak E 2015 {\em Physical Review E\/} {\bf 92}
  052127

\bibitem{borger2008there}
B{\"o}rger L, Dalziel B~D and Fryxell J~M 2008 {\em Ecology Letters\/} {\bf 11}
  637--650

\bibitem{williams1966natural}
Williams G~C 1966 {\em The American Naturalist\/} {\bf 100} 687--690

\bibitem{pnas}
Dannemann T, Boyer D and Miramontes O 2018 {\em Proceedings of the National
  Academy of Sciences\/} {\bf 115} 3794--3799

\bibitem{LotkaElements}
Lotka A~J 1956 {\em Elements of physical biology\/} (Dover Publications)

\bibitem{VolterraVariations}
Volterra V 1928 {\em ICES Journal of Marine Science\/} {\bf 3} 3--51

\bibitem{TauberPopul}
T$\ddot{\rm a}$uber U~C 2012 {\em Journal of Physics A: Mathematical and
  Theoretical\/} {\bf 45} 405002

\bibitem{spencer1990operationally}
Spencer S~R, Cameron G~N and Swihart R~K 1990 {\em Ecology\/} {\bf 71}
  1817--1822

\bibitem{foerster2002home}
Foerster C~R and Vaughan C 2002 {\em Biotropica\/} {\bf 34} 423--437

\bibitem{weimerskirch2007seabirds}
Weimerskirch H 2007 {\em Deep Sea Research Part II: Topical Studies in
  Oceanography\/} {\bf 54} 211--223

\bibitem{wood2017home}
Wood L~D, Brunnick B and Milton S~L 2017 {\em Journal of Herpetology\/} {\bf
  51} 58--67

\bibitem{camposmendez2015}
Campos D and M\'endez V~m~c 2015 {\em Phys. Rev. E\/} {\bf 92}(6) 062115

\bibitem{viswanathan1999optimizing}
Viswanathan G~M, Buldyrev S~V, Havlin S, Da~Luz M, Raposo E and Stanley H~E
  1999 {\em Nature\/} {\bf 401} 911--914

\bibitem{bartulevin}
Bartumeus F and Levin S~A 2008 {\em Proceedings of the National Academy of
  Sciences\/} {\bf 105} 19072--19077

\end{thebibliography}

\providecommand{\newblock}{}

\end{document}